\documentclass[seceq,preprint]{ptptex}

\usepackage{graphicx}
\newcommand{\be}{\begin{equation}}
\newcommand{\ee}{\end{equation}}
\newcommand{\bea}{\begin{eqnarray}}
\newcommand{\eea}{\end{eqnarray}}
\newcommand{\beann}{\begin{eqnarray*}}
\newcommand{\eeann}{\end{eqnarray*}}
\newcommand{\nn}{\nonumber}
\newcommand{\ba}{\begin{array}}
\newcommand{\ea}{\end{array}}

\newcommand{\Tr}{{\rm Tr}\,}
\newcommand{\diag}{{\rm diag}\,}
\newcommand{\Sdet}{{\rm Sdet}\,}
\newcommand{\e}{\epsilon}
\newcommand{\et}{\tilde{\epsilon}}
\newcommand{\Z}{\mathbb{Z}}
\newcommand{\R}{\mathbb{R}}
\newcommand{\C}{\mathbb{C}}

\newcommand{\del}{\partial}

\newcommand{\lambdat}{\tilde{\lambda}}
\newcommand{\Phib}{\bar{\Phi}}
\newcommand{\nv}{{\bf n}}
\newcommand{\iv}{{\bf i}}
\newcommand{\jv}{{\bf j}}
\newcommand{\N}{\mathbb{N}}
\newcommand{\vars}{\varsigma}

\title{
Lattice Formulation\\of\\Two Dimensional Topological Field Theory
}

\author{%       %Use \scshape  for the family name
Kazutoshi \textsc{Ohta}$^{1,}$\footnote{E-mail: kohta@phys.tohoku.ac.jp} 
and Tomohisa \textsc{Takimi}$^{2,}$\footnote{E-mail: ttakimi@riken.jp}
}

\inst{%         %Affiliation, neglected when [addenda] or [errata]
{\small
$^1$Department of Physics, 
Graduate School of Science,
Tohoku University,\\
Sendai 980-8578, Japan
\\
$^2$Theoretical Physics Laboratory,\\
The Institute of Physical and Chemical Research (RIKEN),\\
Wako, Saitama 351-0198, Japan\\
and\\
Yukawa Institute for Theoretical Physics, Kyoto University,\\ Kyoto 606-8502, Japan
}
}

\abst{
We investigate an integrable property and observables of 
2 dimensional ${\cal N} =(4,4)$ topological field theory 
defined on a discrete lattice
by using the ``orbifolding'' and ``deconstruction'' methods. 
We show that our lattice model possesses the integrability
and the partition function reduces to matrix integrals of scalar fields on sites
in consequence.
We make clear meaningful differences 
between the discrete lattice and differentiable manifold,
which would be important to a study of  topological quantities on the lattice.
We also propose a new construction of ${\cal N} =(2,2)$
supersymmetric lattice theory, which is realized by a suitable
truncation of scalar fields from the ${\cal N} =(4,4)$ theory.
}

\preprintnumber{
hep-lat/0611011 \\
RIKEN-TH-90\\
TU-779\\
YITP-06-56}

\notypesetlogo

\begin{document}

\maketitle

\section{Introduction}

Non-perturbative dynamics in gauge and string theory are 
very important issues.
%but there is no efficient formulation in investigating exactly.
A lattice formulation of gauge theory is one of candidate tools
to analyze non-perturbatively in strong coupling region,
but the whole picture including non-trivial topological configurations
like instantons
still has not been known.

On the other hand, some non-perturbative dynamics can be found exactly
in continuum gauge field theory
if the theory possesses supersymmetry.
Especially, if there exist the sufficient numbers of preserving supercharges,
holomorphic observables including corrections from many instantons has
been already known exactly. Seiberg and Witten show that a prepotential in
4 dimensional ${\cal N}=2$ supersymmetric gauge theory is exactly
given in terms of a hyper-elliptic curve (Seiberg-Witten curve) which contains
multi instanton corrections in principle \cite{Seiberg:1994rs,Seiberg:1994aj}.
More recently, Nekrasov also shows that the prepotential arises as a free energy
of some statistical partition function, where the author uses a technique of the so-called
``localization'' \cite{Nekrasov:2002qd,Nekrasov:2003af}.
Moreover Dijkgraaf and Vafa show that an effective superpotential
of ${\cal N}=1$ supersymmetric gauge theory can be obtained from a holomorphic
matrix model \cite{Dijkgraaf:2002fc,Dijkgraaf:2002vw,Dijkgraaf:2002dh}.

These exact results are thanks to integrability of the supersymmetric gauge theory,
and the more supercharges the system has, the more conspicuous the integrable nature becomes.
Then we may ask a natural question; if we realize the supersymmetric gauge theories
on the discrete lattice, is the integrable structure still hold? If so, are the various
non-perturbative results, which can be obtained by the algebraic curves and matrix models,
also recovered from lattice theory? To investigate this structure
is our motivation and purpose in the present paper.

Constructions of supersymmetric gauge theories on the lattice are considered in
various ways. 
\cite{Cohen:2003qw}$^-$\cite{Suzuki:2005dx}
Most of constructions concentrate one's attention only on those actions themselves,
which of course coincide with the continuum theory
in the limit that the lattice spacing goes to zero,
but 
there are other several studies \cite{Onogi:2005cz,Giedt:2003ve}$^-$
\cite{Catterall}
beyond a tree level.
%T.Onogi and T.T perform the perturbative studies of \cite{Cohen:2003xe}
%in their paper.
%J.Giedt perform the non-perturbative
%investigation on the supersymmetric lattice theory
%And  
%Elliot and Moore did the perturbative calculation of
%the 3 dimensional ${\cal N}$ = 2 theory.}.
As we said, we would like to explore
the integrability of
lattice theory as well as continuum theory.
We expect that some classes of the observables may be obtained exactly even on the lattice.
Topological field theory is a toy model but good example
to see the integrable structure of field (lattice) theory.
Basic algebraic structure of topological field theory on the lattice is considered
in \citen{Fukuma:1992hy} physically and developed in \citen{Pfeiffer}
from the mathematical point of view.
More explicit construction of topological field theory on the lattice
is first proposed in \citen{Catterall:2003wd},\citen{Sugino:2003yb}.

To proceed analysis of topological field theory on the lattice, we utilize
the formulation via  ``orbifolding''
and
``deconstruction'', which are discovered by 
\citen{Douglas:1996sw} and \citen{Arkani-Hamed:2001ie},
 and
applied to
supersymmetric lattice theory by \citen{Cohen:2003qw}-\citen{Kaplan:2002wv}. 
This is because these constructions seem to be
natural from a D-brane construction point of view, namely the deconstruction
technique is essentially equivalent to quiver gauge theories or D-branes at an orbifold
singularity, which are very compatible with string theoretical interpretations.
So we think that the deconstruction may preserve the integrable structure (and partial
supersymmetries) even in the discrete lattice picture.

In this paper, we formulate 2 dimensional topological field theory
and investigate characters of the partition function and topological observables
by using the deconstruction and the localization. 
We first construct 2 dimensional
${\cal N}=(4,4)$ from a reduced matrix model with 8 supercharges. And also
we discuss on the properties of the observables of the reduced matrix model,
the orbifolded model and the deconstructed model. 
%We also consider how the observables
%in the continuous theory recover by the deconstruction.
In section \ref{Sec:partition}, 
we derive an exact partition function, 
which coincides with  by the localization technique. 

Organization of the paper is as follows:
In section~\ref{Sec_MM} 
we explain about the reduced matrix model from which we 
construct the supersymmetric lattice formulation.
The reduced model is obtained from a dimensional reduction of 
6 dimensional ${\cal N} = 1$ supersymmetric gauge theory.
There we fix some notations and an original action to use 
in the lattice construction.
And in section~\ref{Sec_LM}, we construct a 2 dimensional 
${\cal N} =(4,4)$ supersymmetric lattice model
from the reduced matrix model using the ``orbifolding'' and 
``deconstruction'' methods. We also 
explain about these methods there.
And in the last part of the section, we show a new construction 
of an ${\cal N} =(2,2)$ supersymmetric lattice formulation by 
the ``truncation'' of scalar fields from the ${\cal N} =(4,4)$ 
lattice theory.
And in section~\ref{Sec_PO} which is a main part of this paper,
we discuss about the partition function and observables 
on the lattice theory.
%And there we discuss about 
%relationship between the observables
%the topology of base manifold.
Section~\ref{Sec_Conc} is devoted to a conclusion and discussions.
Some additional definitions associated with 
a representation of Clifford algebra are denoted in 
appendix~\ref{Notation}.

%\\

%We first give some definitions and notations on the reduced matrix model
%from 6 dimensional ${\cal N}=1$ supersymmetric gauge theory.
%We next perform the orbifolding associated with $\C^2/\Z_N\times \Z_N$,
%which is a seed of the deconstruction to two dimensional theory.

%\section{Introduction}
\section{Reduced Matrix Model with 8 Supercharges}
\label{Sec_MM}
In prior to constructing topological lattice theory,
we start from an explanation of the reduced matrix model with 8 supercharges
since it is the mother of lattice theory arising after the orbifolding and deconstruction. 
In this section, we fix some notations and an original action to use in the following constructions.

In order to construct 
the 0 dimensional supersymmetric matrix model with 8 supercharges,
it is convenient to consider a
dimensional reduction from
${\cal N}=1$
supersymmetric $U(MN^2)$ 
Yang-Mills theory in the Wick-rotated Euclidean 6 dimensional space-time
\be
S=\frac{1}{g^2} \int d^6x \Tr \left\{ \frac{1}{4} F^{KL}F_{KL} 
+\frac{i}{2}\bar{\Psi} \Gamma^{L}D_{L} \Psi \right\},
\label{N=1 6d}
\ee
where $K,L$ are space-time indices and run from $1$ to $6$,
and $\Gamma^K$ satisfy the Euclidean space Clifford algebra
\[
\{\Gamma_K,\Gamma_L\}=2\delta_{KL}.
\] 
Here we take $\Gamma^{K} (K \ne 6) $ as $8\times 8$ symmetric matrices 
and $\Gamma^6$ as antisymmetric.
%In the Euclidean theory, 
%which is obtained by Wick rotation, 
Although the spinor
$\Psi$ 
is defined as an
8 component {\it real} fermion in the 6 dimensional Minkowski space-time, 
here it is interpreted as a {\it complexified} 8 component fermion in the Euclidean space-time.
This interpretation is needed to
have the  well-defined $SO(6)$ rotational symmetry
of the action
since 
the Lorentz boost cannot keep 
the property of $\Psi$ being real
after the Wick rotation \cite{Sugino:2003yb,vanNieuwenhuizen:1996tv}.
%which is 
%enhanced from 8 component real fermion
%composed by 
Here we can impose as $\bar{\Psi}=-i\Psi^{T}\Gamma^6$
with the complexified $\Psi$.
The action (\ref{N=1 6d}) 
is invariant also under the following supersymmetry transformations
\be
\begin{split}
&\delta A_K = \e^T \gamma_K \Psi,\\
&\delta \Psi = -\frac{1}{2}\left( F_{KL}\gamma^{KL}
+2F_{K6} \gamma^K \right)\e,
\label{SUSY transformation}
\end{split}
\ee
where 
$\gamma^K = -\Gamma^6 \Gamma^K$,  
$\gamma_{KL}=\frac{1}{2}[\gamma_K,\gamma_L]$, 
and $\e$ is a Killing spinor.
These $\gamma_K$ are 
introduced
in order to absorb $\Gamma^6$ of $\bar{\Psi}$ into $\Gamma^K$.

The 0 dimensional reduced matrix model is 
obtained by ignoring derivatives in the action
(\ref{N=1 6d})
\be
S_0 = -\frac{1}{g^2}\Tr \left\{
\frac{1}{4} [ A_{K}, A_{L}]^2 
+\frac{i}{2} \Psi^{T} \gamma^{K} [ A_{K}, \Psi] \right\},
\label{reduced MM}
\ee
where $A_K$
are $MN^2\times MN^2$ Hermitian matrices.
%, 
%and we define
%\[
%\gamma^{K} = -i \Gamma^{6}\Gamma^{K},
%\]

Introducing the following complex matrix coordinates
\be
\begin{array}{ll}
X=A_1-iA_4, & X^\dag=A_1+iA_4,\\
Y=iA_2-A_3, & Y^\dag= -iA_2-A_3,\\
\Phi=A_5+iA_6, & \bar{\Phi}=A_5-iA_6,
\end{array} \label{complex-boson-matrix}
\ee
the bosonic part of the action (\ref{reduced MM}) can be written as
\[
\begin{split}
S_0|_B&=\frac{1}{g^2}\Tr\Bigg\{
\frac{1}{8}([X,X^\dag]+[Y,Y^\dag])^2+\frac{1}{2}|[X,Y]|^2\\
&\quad+\frac{1}{4}(|[X,\Phi]|^2+|[X,\Phib]|^2+|[Y,\Phi]|^2+|[Y,\Phib]|^2)\\
&\quad+\frac{1}{8}|[\Phi,\Phib]|^2
\Bigg\}.
\end{split}
\]
Similarly, taking $\Psi$ as the % 'TFT form',
following form by an $8 \times 8$ 
matrix $U_f$,
\[
\Psi^T=(\lambda_2,\lambda_3,\lambda_4,\lambda_1,\chi^{12},\chi^{13},
\chi^{14},\frac{1}{2}\eta)
=(\lambda,\lambda^\dag,\lambdat,\lambdat^\dag,\chi^\R,\chi^\C,
{\chi^\C}^\dag,\frac{1}{2}\eta) \cdot U_f^T,
\]
for a suitable representation of $\gamma^K$,
the fermionic part becomes
\[
S_0|_F=S_{F1}+S_{F2}+S_{F3},
\]
and
\bea
S_{F1}&=& \frac{1}{2g^2}
\Tr \biggl(
-\chi_\C [X^\dag, \lambdat^\dag ]
-(\frac{1}{2}\eta +i\chi_\R ) [Y, \lambdat^\dag ]
+\chi_\C [Y^\dag, \lambda^\dag ]
-(\frac{1}{2}\eta +i\chi_\R ) [X, \lambda^\dag ]
\nonumber\\
& &-(\frac{1}{2}\eta -i \chi_\R ) [X^\dag, \lambda ]
+\chi_\C^\dag [Y, \lambda ]
-(\frac{1}{2}\eta -i\chi_\R) [Y^\dag, \lambdat ]
-\chi_\C^\dag [X, \lambdat ] \biggr),
\nn\\
S_{F2}&=& \frac{1}{4g^2}
 \Tr \biggl(
\lambdat^\dag [\bar{\Phi},\lambdat] 
+\lambdat [\bar{\Phi},\lambdat^\dag ] 
+\lambda [\bar{\Phi},\lambda^\dag ] 
+\lambda^\dag [\bar{\Phi},\lambda ]
\biggr),
\nn\\
S_{F3}&=&
-\frac{1}{2 g^2} \Tr \left(
\frac{1}{2}(\chi_\C [\Phi, \chi_\C^\dag ] 
+\chi_\C^\dag [\Phi, \chi_\C ] )
+\chi_\R [\Phi, \chi_\R ] 
+\frac{1}{2} \eta [ \Phi , \frac{1}{2} \eta ]
\right).\nn
\eea 
See appendix \ref{Notation} for an explicit form of $\gamma^K$ and $U_f$.
The supersymmetric transformations of the reduced matrix model can be obtained from the dimensional reduction
of (\ref{SUSY transformation}). Especially, if we pick up the last component of the Killing spinor
as $\e=(0,\ldots,0,-\varepsilon)$, we can extract a part of the supersymmetric transformations proportional
to a scalar supercharge.
Denoting the scalar supercharge as $Q$, we get the scalar supersymmetric transformations
for the matrix components
\be
\begin{array}{ll}
QX=\lambda, & Q\lambda =[\Phi,X],\\
QY=\lambdat, & Q\lambdat =[\Phi,Y],\\
Q\vec{H}=[\Phi,\vec{\chi}], & Q\vec{\chi} = \vec{H},\\
Q\Phib=\eta, & Q\eta=[\Phi,\Phib],\\
Q\Phi=0, &
\label{BRST-def}
\end{array}
\ee
where $\vec{H}\equiv(H^\R,H^\C,{H^\C}^\dag)$ is introduced as auxiliary matrix variables in the off-shell
supersymmetric transformations.
These transformations make a BRST algebra. So we call the transformations as the BRST transformations.
We notice that almost matrices have supersymmetric (BRST) partners and make pairings, but
only $\Phi$ lives alone without any partner. So if we denote the bosonic degrees of freedom as
\[
\vec{\cal B}=(X,X^\dag,Y,Y^\dag,\vec{H},\Phib),
\]
then the corresponding fermionic matrices are
\[
\vec{\cal F}=(\lambda,\lambda^\dag,\lambdat,\lambdat^\dag,\vec{\chi},\eta).
\]
$\Phi$ is independent of these variables.
Here we define the pair of them as
$\vec{\cal A} \equiv (\vec{ \cal B},\vec{\cal F})$.
Be attention to that the BRST transformations 
%of 
%$\vec{\cal A}$
are written in terms of 
homogeneous transformations of $\vec{\cal A}$.
%And $\Phi$ is BRST closed.

The BRST transformations play a special role in the theory. Indeed, the action (\ref{reduced MM})
can be written as a BRST exact form \cite{Hirano:1997ai,Moore:1998et}
%(whether topological twisted theory or not)
\be
S_0 = \frac{1}{2g^2}Q \Xi(\vec{\cal B},\vec{\cal F},\Phi),
\label{BRST exact action}
\ee
where
\[
\Xi(\vec{\cal B},\vec{\cal F},\Phi) = \Tr\left[ \frac{1}{4}\eta[\Phi,\Phib]
+\vec{\chi}\cdot(\vec{H}-i\vec{{\cal E}})
+\frac{1}{2}\left(\lambda[X^\dag,\Phib]+\lambda^\dag[X,\Phib]
+\lambdat[Y^\dag,\Phib]+\lambdat^\dag[Y,\Phib]\right)\right].
\]
Here $\vec{\cal E} \equiv ({\cal E}^\R,{\cal E}^\C,{{\cal E}^\C}^\dag)$ are defined as
\bea
{\cal E}^\R&=&-([X,X^\dag]+[Y,Y^\dag]),\nn\\
{\cal E}^\C&=&2i[X,Y].\nn
\eea
They
are associated with the D- and F-term conditions in 4 dimensional ${\cal N}=2$ supersymmetric gauge theory
and constrained to be zero by the Lagrange multiplier matrices $\vec{H}$.
These constrained conditions correspond to the ADHM equations for the self-dual field strength (instantons)
 in 4 dimensional gauge theory.

Now let us consider the global symmetry of the reduced matrix model,
which is important to the \textit{topological twisting} and the
\textit{orbifolding} of the theory.
%which is denoted in the next section.
%The global symmetry of the theory 
%is $SO(6)\times U(1)$, which are the rotational symmetry
%and $R$-symmetry, respectively.
%By the argument of CKKU\cite{Cohen:2003qw}
First of all, this reduced model from 
6 dimensional ${\cal N} =1$ supersymmetric Yang-Mills theory has 
$G = SO(6) \times SU(2)_I$ global symmetry, where
$SO(6)$ is a rotational symmetry of the original 6 dimensional theory
and $SU(2)_I$ is an enhanced $R$-symmetry emerged
when $\Psi$ is complexified.
Moreover the symmetry includes a subgroup as like as
\be
G=SO(6)\times SU(2)_I \supset SU(2)_L \times SU(2)_R \times SU(2)_I,
\label{global symmetry}
\ee
where $SU(2)_L\times SU(2)_R$ is isomorphic to the $SO(4)$ rotational symmetry 
in 4 dimensions.
%we take 8 real fermions as complex.
%($\lambda^\dag$ is 
%not complex conjugate of 
%$\lambda$, 
%independent of 
%$\lambda$. 
%So from here, this theory has eight complex fermion fields) 
%of ${\cal N}=2$ theory.
We will utilize this partial global symmetry to construct 
a topologically twisted theory.
%\\
%First of all, 
We now introduce a quaternionic $2\times 2$ matrix
notation of the 4 dimensional bosonic matrix coordinates  
\be
{\bf X}\equiv
%X_\mu\bar{\sigma}^\mu
%=
\begin{pmatrix}
Y & X\\
-X^\dag & Y^\dag
\end{pmatrix},
%(\mu = 1, \ldots 4 ),
\label{X1}
\ee
and also %a $2\times 2$ matrix notation
 for the fermionic matrix coordinates, 
\be
{\bf \Lambda}\equiv
\begin{pmatrix}
\lambdat^\dag & -\lambda\\
\lambda^\dag & \lambdat
\end{pmatrix},\qquad
\bar{\bf \Lambda}\equiv
\begin{pmatrix}
\frac{1}{2}\eta+i\chi^\R & -\chi^\C\\
{\chi^\C}^\dag & \frac{1}{2}\eta-i\chi^\R
\end{pmatrix}.
\label{L1}
\ee
These transform as
\bea
&&{\bf X} \rightarrow R{\bf X}L^\dag, \label{trans-b}\\ 
%\ee
%\bea
&&{\bf \Lambda} \rightarrow L{\bf \Lambda}M^\dag, \label{trans-f}\\ 
&&\bar{\bf \Lambda} \rightarrow M\bar{\bf \Lambda}R^\dag,\label{trans-f2} 
\eea
where $L\in SU(2)_L$ , $R\in SU(2)_R$ and
$M\in SU(2)_I$.% belongs to the enhanced internal $R$-symmetry. 
%and $U(MN^2)$ part is identical.
%The $SU(2)_L\times SU(2)_R$ Lorentz symmetry acts on the above matrix as

The bosonic part of the action $S_0|_B$ is 
originally invariant under the 
$SO(6) \supset SU(2)_L \times SU(2)_R$ Lorentz symmetry and 
trivial under $SU(2)_I$.
Using the quaternionic notation of matrix variables
(\ref{X1}) and (\ref{L1}),
each fermionic part 
$S_0|_F = S_{F1} +  S_{F2} + S_{F3}$ 
can be written as
\bea
&&S_{F1} = -\frac{1}{2g^2} \Tr \bar{\bf \Lambda} [ {\bf X}, {\bf \Lambda}],
\nn\\
&&S_{F2} = \frac{1}{4g^2}\Tr {\bf \Lambda} 
[ \Phib \otimes {\bf 1}_2, \sigma_2{\bf \Lambda}^T\sigma_2], 
\nn\\
&&S_{F3} = -\frac{1}{4g^2} \Tr \bar{\bf \Lambda}
[ \Phi \otimes {\bf 1}_2, \sigma_2\bar{\bf \Lambda}^T\sigma_2], \nn
\eea
where $\sigma_2$  acts only on the $2\times 2$ matrix 
indices of the quaternionic notation %${\bf X},{\bf \Lambda},\bar{\bf \Lambda}$.
and the commutator stands only for the $MN^2$ gauge indices.
%Bethe commutation affects 
From these equations,
we can see immediately $S_{F1}$ is  
invariant under the 
transformation
(\ref{trans-b})--(\ref{trans-f2}).
The invariance of $S_{F2}$ and $S_{F3}$ is also found
by regarding $\sigma_2{\bf \Lambda}^T\sigma_2$ or
$\sigma_2\bar{\bf \Lambda}^T\sigma_2$
as the complex conjugate representation of ${\bf \Lambda}$ or
$\bar{\bf \Lambda}$.
%especially 

The topological twisting amounts to a redefinition of the global symmetry
by taking a diagonal part $SU(2)'$ of $SU(2)_R\times SU(2)_I$.
This means that we need to set $M=R$. Under the redefined symmetry
$SU(2)_L\times SU(2)'$, BRST charge $Q$ becomes a singlet
and ${\bf X}^\dag$ and ${\bf \Lambda}$ transform
as the same way,
where ${\bf X}^\dag$ is given by
\be
{\bf X}^\dag =
\begin{pmatrix}
Y^\dag & -X\\
X^\dag & Y
\end{pmatrix},\label{Xd1}
\ee
and transforms as 
\be
{\bf X}^\dag \rightarrow L{\bf X}^\dag R^\dag.
\label{trans-bd}
\ee

Thus we can identified all global symmetries which can be used for the orbifolding on the reduced matrix model.

\section{Two Dimensional Lattice Theory from Topological Matrix Model}
\label{Sec_LM}
%\section{Deconstruction of Two Dimensional Field Theory}

In this section we explain how we can obtain the lattice formulation 
which preserves BRST charges. 
%and possesses property of the decent relation 
%and the Nicolai map.
We can construct such lattice formulation 
%of two dimensional topological field theory
%which can preserve same 
%BRST charge 
by the orbifolding and the deconstruction methods.
\cite{Cohen:2003qw,Cohen:2003xe,Kaplan:2002wv}

The orbifolding is the projection
of matrix variables to the invariant subspace under 
$\Z_N \times \Z_N \subset G
\times U(MN^2)$, where $G \equiv  SU(2)_L \times SU(2)_R \times SU(2)_I$.
An appropriate choice of generator sets of $\Z_N \times \Z_N$ lets us 
construct an orbifold action which preserves fermionic supercharge 
which is equivalent to the BRST charge of topological field theory.
Theory orbifolded by the $\Z_N \times \Z_N$ is also regarded as the
reduced matrix model of quiver gauge theory \cite{Douglas:1996sw}
or brane box model \cite{Hanany:1997tb,Hanany:1998it}.

We have to perform not only the orbifolding but also the deconstruction
to realize lattice gauge theory in 
the 2 dimensional space-time. 
%from the zero dimensional 
%matrix theory
The orbifolded quiver action itself cannot be regarded as a lattice action % with space-time
since these have no kinetic term.
A mechanism called as deconstruction automatically
generates kinetic terms by a spontaneous breakdown of the gauge symmetry.
By the deconstruction,
the bosonic link fields $X,X^\dag,Y,Y^\dag$ are redefined as a fluctuation around the vacuum
expectation value which is characterized as $1/a \times {\bf 1}_M$,
where $a$ stands for a lattice spacing.
%This vacuum expectation value is exactly same as 
%back ground matrices
%(\ref{backx})-(\ref{backy}).
%This redefinition automatically generates the kinetic terms.
%We first obtain desired lattice formulation by performing
%orbifolding and deconstruction.

We perform this orbifolding and deconstruction procedure step by step in the following.

\subsection{Orbifolding}

We first consider orbifold projection operators 
$\hat{\gamma}_a \in \Z_N \times \Z_N$ ($a=1,2$)
acting on an $MN^2 \times MN^2$ matrix variable
$\tilde{O}$ as
%are defined as follows
\bea
&(\hat{\gamma}_a\tilde{O})_{m,n} \equiv 
(e^{i2\pi r_a/N}
\mathcal{C}_{(a)}\tilde{O}\mathcal{C}_{(a)}^{-1}
)_{m,n}, \label{orb-def} \\
&e^{i2\pi r_a/N} \in %\Z_N \times \Z_N  \subset
G, %\equiv  SU(2)_L \times SU(2)_R \times SU(2)_I,
\qquad \mathcal{C}_{(a)}
\in %\Z_N \times \Z_N
U(MN^2),\nn
\eea
where indices $m,n$ are $U(MN^2)$ gauge indices of an 
adjoint representation.
%and 
%$e^{i2\pi r_a/N} \in \Z_N \times \Z_N  \subset G \equiv 
%SU(2)_L \times SU(2)_R \times SU(2)_I$,  $\mathcal{C}_{(a)}
%\in \Z_N \times \Z_N
%\subset U(MN^2)$.
%And here we represent $\tilde{O}$
%as general matrix of 
%matrix theory.

The orbifold projection mods out
matrix components which do not satisfy
\[
(\hat{\gamma}_a\tilde{O})_{m,n} = \tilde{O}_{m,n} 
\quad \text{for any } a.
\]
Due to this projection, the gauge symmetry $U(MN^2)$ breaks down to 
$U(M)^{N^2}$.
The supercharges in the matrix model 
are also projected out
since they transform under $G$.
Projection conditions for the supercharges are 
defined as
\be
(\hat{\gamma}_a Q)_{\vars} = (e^{i2\pi r_a/N}Q)_{\vars} =
Q_{\vars} \quad \text{for any } a,
\label{SUSY-orb}
\ee
where we denote $\vars$ as indices of 
$G$
 and $\mathcal{C}_{(a)}$ are canceled out 
since 
the supercharges are gauge singlets.
The supercharges which survive under this condition 
become preserved fermionic charges on the lattice.

We define explicitly generators $\mathcal{C}_{(a)}$ from the gauge group as  
\be
\begin{array}{ll}
\mathcal{C}_{(1)} = \Omega \otimes {\bf 1}_N \otimes {\bf 1}_M, & \\
\mathcal{C}_{(2)} = {\bf 1}_N \otimes \Omega \otimes {\bf 1}_M, & \\
\Omega = \begin{pmatrix}
	\omega & & & & \\
	 & \omega^2 &  &  &  \\
	 &  & \ddots & &  \\
	 &  &  & \omega^{N-1} &  \\
	 &  &  &  & 1
	\end{pmatrix},
& \omega=e^{2\pi i/N}. 
\end{array}
\label{C-def}
\ee
Then $e^{i2\pi r_a/N}$ are combinations of the Cartan subgroup (maximal torus)
of the global symmetry  $G$. %and $r_a$ are linear combinations of those $U(1)$ charges.
We should define the $r_a$ charges 
to make the supercharge which is equivalent to
the BRST charge surviving under the projection
(\ref{SUSY-orb}).
We will define the set of $r_a$ charges using by the Cartan
subgroup of $SU(2)_L \times SU(2)'$
since the BRST charge is invariant under the $SU(2)_L \times SU(2)'$.
If we denote $U(1)$ charges of each Cartan subgroup of $SU(2)$ factor in 
$SU(2)_L \times SU(2)_R \times SU(2)_I$ 
%(\ref{global symmetry})
as $L_3$, $R_3$ and $M_3$, respectively,
%the BRST charge is invariant under the $SU(2)_L \times SU(2)'$,
the corresponding $U(1)$ charges 
of $SU(2)_L \times SU(2)'$
are $L_3$ and $R_3' \equiv R_3 + M_3$. 
% \, (i = 1,2,3)$.
%where $R_i$ and $M_i$ are generators of $SU(2)_R$ and $SU(2)_I$ respectively,
%and Cartan subalgebra of  $su(2)_{diag}$ is $R_3' = R_3+M_3$.
So we define the set of $r_a$ charges as linear combinations of $L_3$ and 
$R_3'$
\be
\begin{split}
r_1 &\equiv L_3+R_3' = L_3+R_3+M_3,\\
r_2 &\equiv -L_3+R_3' = -L_3+R_3+M_3.
\end{split}
\label{r-charge-def}
\ee
When we perform the orbifold projection
with respect to these charges, 
we can preserve the BRST charge on the lattice.
Besides, under these $U(1)$ symmetries, 
the matrix variables $X,Y,\lambda, {\it etc}.$ become eigenmatrices
with eigenvalues denoted in Table \ref{U(1) charges}.

\begin{table}[h]
\begin{center}
\begin{tabular}{|c||c|c|c|c|c|c|}
\hline
 & $X,\lambda$ & $Y,\lambdat$ & $H^\R,\chi^\R$ & $H^\C,\chi^\C$ & $\Phib,\eta$ & $\Phi$ \\
 \hline
$r_1$ & $+1$ & $0$ & $0$ & $+1$ & $0$ & $0$ \\
$r_2$ & $0$ & $+1$ & $0$ & $+1$ & $0$ & $0$ \\
\hline
\end{tabular}
\end{center}
\caption{$U(1)$ charges}
\label{U(1) charges}
\end{table}

From the definition (\ref{orb-def}), (\ref{C-def}) and (\ref{r-charge-def}),
each $MN^2 \times MN^2$ matrix variable reduce to the $N^2$ sets of 
$M \times M$ submatrices by the orbifold projection if we decompose each matrix 
into $N^2 \times N^2$ blocks of $M \times M$ submatrices.
The $N^2$ products of $U(M)$ gauge symmetry acts on these $M\times M$ submatrices.
Then we label the $N^2$ 
%\times N^2$ 
row indices for different blocks as $\nv=(n_1,n_2)$ 
($n_{1,2}=1,\ldots,N$), which will become
coordinates (sites) of 
2 dimensional space-time (lattice). % and 
%we obtain the quiver lattice structure which preserve BRST charge.

\begin{figure}[t]
\begin{center}
\includegraphics[scale=0.65]{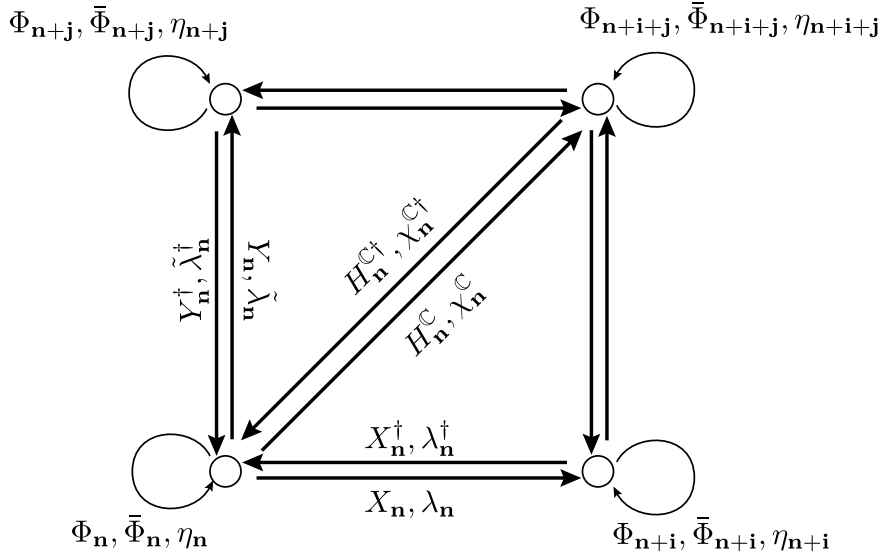}
\end{center}
\caption{A quiver diagram for the orbifold model.}
\label{fig-lat}
\end{figure}

We now define $N^2$ sets of the vector $\vec{\cal A}_\nv\equiv(\vec{\cal B}_\nv,\vec{\cal F}_\nv)$
which composed by the $M\times M$ bosonic and fermionic matrix fields respectively
as follows
\[
\vec{\cal B}_\nv=(X_\nv,X_\nv^\dag,Y_\nv,Y_\nv^\dag,\vec{H}_\nv,\Phib_\nv),\qquad
\vec{\cal F}_\nv=(\lambda_\nv,\lambda_\nv^\dag,\lambdat_\nv,\lambdat_\nv^\dag,\vec{\chi}_\nv,\eta_\nv).
\]
We also define the $N^2$ matrices $\Phi_\nv$ which is from
$\Phi$.
Using this manipulation,
we can obtain the following quiver action from the original matrix model 
(\ref{reduced MM})
\be
S=\frac{1}{g^2}\sum_\nv Q\Xi(\vec{\cal B}_\nv,\vec{\cal F}_\nv,\Phi_\nv),
\label{quiver action}
\ee
where
\be
\begin{split}
\Xi(\vec{\cal B}_\nv,\vec{\cal F}_\nv,\Phi_\nv)
&= \Tr\Bigg[\frac{1}{4}\eta_{ {\bf n}}[\Phi_{ {\bf n}},\Phib_{ {\bf n}}]
+\vec{\chi}_\nv\cdot
(\vec{H}_\nv-i\vec{\cal E}_\nv)
\\
&\qquad+\frac{1}{2}
\biggl\{
\lambda_\nv(X_{\nv}^\dag\Phib_\nv-\Phib_{\nv+\iv}X_{\nv}^\dag)
+\lambda_{\nv-\iv}^\dag(X_{\nv-\iv}\Phib_\nv-\Phib_{\nv-\iv}X_{\nv-\iv})\\
&\qquad
+\lambdat_\nv(Y_{\nv}^\dag\Phib_\nv-\Phib_{\nv+\jv}Y_{\nv}^\dag)
+\lambdat_{\nv-\jv}^\dag(Y_{\nv-\jv}\Phib_\nv-\Phib_{\nv-\jv}Y_{\nv-\jv})
\biggr\}\Bigg],
\end{split}
%\label{orb-act}
\ee
and
\bea
{\cal E}^\R_\nv &=& -(X_\nv X_\nv^\dag
-X_{\nv-\iv}^\dag X_{\nv-\iv}
+Y_\nv Y_{\nv}^\dag
-Y_{\nv-\jv}^\dag Y_{\nv-\jv}),\nn\\
{\cal E}^\C_\nv&=&2i(X_\nv Y_{\nv+\iv}-
Y_\nv X_{\nv+\jv}).\nn
\eea
The summation in the action (\ref{quiver action}) is taken over the $N^2$ labels $\nv$,
and 
we define
$\iv$ and $\jv$ as unit vectors in each direction, 
namely $\iv=(1,0)$ and $\jv=(0,1)$, respectively.
The matrices transform as bi-fundamental or adjoint representation under the broken gauge symmetry $U(M)^{N^2}\subset U(MN^2)$.
We represent
the gauge transformation of these matrices by using the quiver diagram in Fig.~\ref{fig-lat}.
From the diagram, we find that
the bosonic $X_\nv,X_\nv^\dag$ ($Y_\nv,Y_\nv^\dag$)
and fermionic $\lambda_\nv,\lambda_\nv^\dag$ ($\lambdat_\nv,\lambdat_\nv^\dag$) fields
%which has each r-charge 
%$(r_1,r_2) = (1,0),(-1,0),(0,1),(0,-1)$ 
reside on links in the $\iv$-  ($\jv$-) direction, 
the fields $\Phi_\nv,\bar{\Phi}_\nv,\eta_\nv,\chi_\nv^\R, H_\nv^\R$ 
%with $r=0$ 
%charge are 
sit on the sites,
and $\chi_\nv^\C, \chi_\nv^{\C\dag}, H_\nv^\C, H_\nv^{\C\dag}$
are diagonal link fields in the quiver diagram. (See Fig.~\ref{fig-lat}.)
The location of each variable is completely decided
by their corresponding $r$-charges. 
%$ex.)$ 
For example,
fields with $(r_1,r_2)=(0,0)$ live at the sites,
ones with $(r_1,r_2)=(1,0)$ reside on links pointing in
the $\iv$-direction, and ones with $(r_1,r_2)=(1,1) $ are 
diagonal link fields.

%%%%%%%%%%%%%%%%%%%%%%%%%%%% Figure%%%%%%%%%%%%%%%%%%%%%%%%%%% Figure
%
%
%%%%%%%%%%%%%%%%%%%%%%%%%%%% Figure%%%%%%%%%%%%%%%%%%%%%%%%%%% Figure
%%%%%%%%%%%%%%%%%%%%%%%%%%% Figure%%%%%%%%%%%%%%%%%%%%%%%%%%% Figure
%%%%%%%%%%%%%%%%%%%%%%%%%%% Figure%%%%%%%%%%%%%%%%%%%%%%%%%%% Figure

By performing the orbifolding to the BRST transformations 
in the matrix model, 
BRST transformations on the lattice are
obtained as follows
\be
\begin{array}{ll}
QX_\nv=\lambda_\nv, & Q\lambda_\nv =\Phi_\nv X_\nv - X_\nv\Phi_{\nv+\iv},\\
QY_\nv=\lambdat_\nv, & Q\lambdat_\nv =\Phi_\nv Y_\nv - Y_\nv\Phi_{\nv+\jv},\\
QH^\R_\nv =[\Phi_\nv,\chi^\R_\nv], & Q\chi^\R_\nv = H^\R_\nv,\\
QH^\C_\nv =\Phi_\nv\chi^\C_\nv-\chi^\C_\nv\Phi_{\nv+\iv+\jv}, & Q\chi^\C_\nv = H^\C_\nv,\\
Q\Phib_\nv=\eta_\nv, & Q\eta_\nv=[\Phi_\nv,\Phib_\nv],\\
Q\Phi_\nv=0. & 
\end{array}\label{orb-BRST}
\ee
These are still homogeneous transformations of $\vec{\cal A}_\nv =
(\vec{\cal B}_\nv, \vec{\cal F}_\nv)$. 

% not 
%only singlet part of $SU(2)_{diag}$.
%So
We would like to comment that
this quiver matrix theory has another preserved charge $Q'$
which is not singlet under the $SU(2)_R \times SU(2)_I$ twisting
but belongs to triplet of the decomposition ${\bf 2}_R \otimes {\bf 2}_I
= {\bf 3} \oplus {\bf 1}$.
The preserved charge $Q'$
is a neutral state in the triplet.
The above orbifolding picks up 
zero-eigenstates of the Cartan 
subalgebra $R_3+M_3$ of $SU(2)'$ including even in the triplet.
However we 
%do not respect to such the triplet charge 
%but 
respect to 
the singlet supercharge
which is used in the topological twisting in the following
since we are interested in a recovering of topological field theory.

%The ordinal BRST charge which is singlet under $SU(2)'$ is 
%one of them,
%include 
%.
%.
%which
%.
%This charge 
%and the $Q'$ is from the triplet part of ${\bf 2_R} \otimes {\bf 2_I}
%= {\bf 3} \oplus {\bf 1}$
%which is 
%generated by the $SU(2)_R \times SU(2)_I$ twisting.
%Method of orbifold projection 
%allows lattice theory to possess this charge
%This is because
%because this method characterize the preserved charge by the 
%0-eigenstates fermion under Cartan subalgebra not by the 
%dimension of irreducible representation.

%%%%%%%%%%%%%%%%%%%%%%%%%%%%%%%%%%%%%%%%%%%%%%%%%%%%%%%%%%%%%%%%%%
\subsection{Geometrical interpretation}

The orbifold model we have constructed has a very nice geometrical interpretation in string theory.
%The quiver gauge theory often appears in the context of effective theory on D-brane in superstring theory.
As well known, open strings on the D-branes produce gauge degrees of freedom and supersymmetric gauge
theory appears as low energy effective theory.
The quiver gauge theory can be realized by
putting the D-branes at an associated orbifold singularity.
If we would like to preserve some of supercharges, the discrete group corresponding to the orbifold
is restricted and classified by a discrete subgroup of a holonomy group in the geometry.
For example, when we orbifold a 2 dimensional complex plane $\C^2$, the quotient group must belong to
the discrete subgroup of $SU(2)$, which is completely classified by ADE Lie algebra.
$SU(2)$ is a holonomy of 4 dimensional hyper-K\"ahler manifold, which preserves the same number of
the supercharges as the orbifold. The A-type orbifold is simply $\C^2/\Z_k$.

%This is nothing but the $\Omega$-background charges.

Let us see the orbifold action of our model on the matrix coordinates in detail.
We first find that the $U(1)$ actions associated with $r_1$ and $r_2$ charges are
\bea
U(1)_1 &:& (X,Y,{H^\C}^\dag)\rightarrow (\omega_1X,Y,\omega_1^{-1}{H^\C}^\dag),\\
U(1)_2 &:& (X,Y,{H^\C}^\dag)\rightarrow (X,\omega_2Y,\omega_2^{-1}{H^\C}^\dag),
\label{U(1) action}
\eea
as denoted in Table~\ref{U(1) charges}.
So the choice of  $\omega_{1,2}=e^{\frac{2\pi i}{N}}$ gives the orbifold $\C^3/\Z_N\times \Z_N$
if we regard
$(X,Y,{H^\C}^\dag)$ as the 3 dimensional complex coordinates.
The discrete group $\Z_N\times \Z_N$ belongs to a discrete subgroup of $SU(3)$. Since $SU(3)$ is a holonomy
of a Calabi-Yau manifold, we expect that the orbifolding preserves some of supersymmetries.

\begin{figure}[t]
\begin{center}
\includegraphics[scale=0.5]{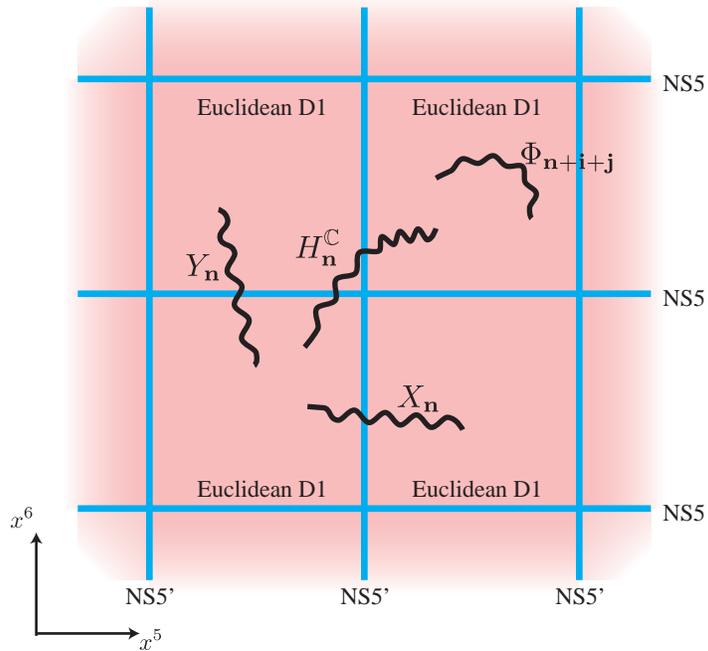}
\end{center}
\caption{The brane configuration of the brane box model. $M$ Euclidean D1-branes expand inside the box
framed by two types of NS5-branes. The total number of boxes (sites) is $N^2$. The open strings connecting between
each neighbor box give the bi-fundamental matrices like $(X,Y,{H^\C}^\dag)$. The brane box is a realization of the
quiver diagram in string theory.}
\label{brane box}
\end{figure}

This orbifold is also considered in the context of a brane configuration, which is a dual
of the orbifold and called as \textit{the brane box model}
\cite{Hanany:1997tb,Hanany:1998it}. The brane configuration of the brane box model consists of D-branes
stretching between NS5-branes. The original brane box model gives 4 dimensional supersymmetric gauge theory
with bi-fundamental hypermultiplets, but the orbifolded quiver model from the reduced matrix model is a 0 dimensional
theory essentially. So world-volumes of branes are spreading as follows
\[
\begin{array}{lccccccccccc}
D1 & : &  &  &  &  &  & 5 & 6 &  &  &  \\
NS5 & : & 0 & 1 & 2 & 3 & 4 & 5 &  &  &  &  \\
NS5' & : & 0 & 1 & 2 & 3 &  &  & 6 & 7 &  & 
\end{array}
\] 
where the numbers stand for the world-volume direction of the branes.
Note that the D1-branes do not extend in the time direction. So the D1-branes are the Euclidean D-branes, whose world-volume
effective theory is topologically twisted \cite{Bershadsky:1995qy}.
We depict the brane box configuration in Fig.~\ref{brane box}.

%%%%%%%%%%%%%%%%%%%%%%%%%%%%%%%%%%%%%%%%%%%%%%%%%%%%%%%%%%%%

%\subsection{Lattice formulation}
\subsection{Deconstruction}

The quiver action 
(\ref{quiver action})
does not include kinetic terms.
To generate kinetic terms, we should 
perform the deconstruction.
The deconstruction is the field redefinition
of the complex bosonic fields
$X_\nv,X_\nv^\dag,Y_\nv,Y_\nv^\dag$ expanding around
vacuum expectation values of
\[
\langle X_\nv \rangle
=\langle Y_\nv \rangle =\frac{1}{a}{\bf 1}_M, 
\quad \langle \Phi_\nv \rangle = 0.
\]
We will interpret $a$ here as the lattice spacing.
%Depending upon discrete theories, 
We can choose two different types of the fluctuations:
One of them is a Cartesian decomposition which is a sum of hermite and antihermite matrices
adopted in
\citen{Cohen:2003qw,Cohen:2003xe,Kaplan:2002wv}. 
Following this decomposition,
we can rewrite the general $M\times M$ complex matrices $X_\nv,Y_\nv$ 
in terms of the hermite matrices $s_{k,\nv}$ ($k=0,\ldots,3$) and $v_{x,\nv},v_{y,\nv}$ as
\[
\begin{array}{ll}
X_\nv=\frac{1}{a}{\bf 1}_M+s_{0,\nv}+iv_{x,\nv}, &
X^\dag_\nv=\frac{1}{a}{\bf 1}_M+s_{0,\nv}-iv_{x,\nv},\\
Y_\nv=\frac{1}{a}{\bf 1}_M+s_{3,\nv}+iv_{y,\nv}, &
Y^\dag_\nv=\frac{1}{a}{\bf 1}_M+s_{3,\nv}-iv_{y,\nv},\\
\Phi_\nv = 
-s_{1,\nv}+is_{2,\nv},&
\end{array}
\]
where $s_{k\nv}$ can be regarded as scalar fields and 
$v_{x\nv}$, $v_{y\nv}$ are gauge fields  
in $x$, $y$-direction,
which are deconstructed 2 dimensional space-time
directions
corresponding to $\iv$, $\jv$-directions in the orbifolded quiver model. 
%This choice is 
If we take this choice, 
the $U(M)$ gauge symmetry is 
obscured on the lattice.
%in this choice.
On the other hand,
another choice proposed in \citen{Unsal:2005yh,Onogi:2005cz}
makes the gauge symmetry manifest on the lattice.
It is a polar decomposition of complex matrices
which is uniquely represented as a product of hermite
matrices $(\frac{1}{a} + s_{k,\nv})$ ($k =0,3$),
which represent a radial direction and so have positive eigenvalues only, and unitary matrices
\be
\begin{array}{ll}
X_\nv=\left(\frac{1}{a}{\bf 1}_M+s_{0,\nv}\right)U_{x,\nv},&
X^\dag_\nv=U_{x,\nv}^\dag\left(\frac{1}{a}{\bf 1}_M+s_{0,\nv}\right),\\
Y_\nv=\left(\frac{1}{a}{\bf 1}_M+s_{3,\nv}\right)U_{y,\nv},&
Y^\dag_\nv=U_{y,\nv}^\dag\left(\frac{1}{a}{\bf 1}_M+s_{3,\nv}\right),\\
\Phi_\nv = -s_{1,\nv}+is_{2,\nv}, &
\end{array}
\label{unitaryxy}
\ee
where $U_{\mu,\nv}$ are unitary matrices
written by using the gauge fields as
$U_{\mu,\nv} = e^{iav_{\mu,\nv}}$ ($\mu = x, y$). 
%And we regard $s_{k} (k= 0,1,2,3)$ are scalar fields
%and $v_x,v_y$ are gauge fields each corresponds to
%one of $x$ and $y$ direction.
Note that the fluctuation must satisfy $|s_{k,\nv}|\ll \frac{1}{a}$
because of the positivity of the radial coordinate matrices 
$(\frac{1}{a} + s_{k,\nv})$.
In this choice, the gauge symmetry are manifest, that is,
the gauge fields are described by compact link fields and the scalar fields 
are sitting on the sites.
%This is 
%This choice is right if the complex matrices is 
If we expand the compact link fields 
as
%\be
$U_{\mu,\nv } = 1+ a v_{\mu,\nv} + {\cal O}(a^2)$,
the polar decomposition reduces to the Cartesian decomposition at 
$a\to 0$ limit.
So we find that the $a\to 0$ limit of the 
both choices give the same continuum theory,
but we adopt the polar
coordinate decomposition in the following constructions
because of the advantage of the manifest gauge symmetry.

One can immediately see the generation of the kinetic term 
on the lattice by substituting the (\ref{unitaryxy})
into the orbifold action (\ref{quiver action}).
For example,
let us consider a part of the action (\ref{quiver action})
\[
 \frac{1}{2g^2}\sum_{\nv}Q\Tr\left[\lambda_\nv(X_{\nv}^\dag\Phib_\nv-\Phib_{\nv+\iv}X_{\nv}^\dag)\right]=
-\frac{1}{2g^2}\sum_\nv
\Tr\left[
\lambda_\nv(X_{\nv}^\dag\eta_\nv-\eta_{\nv+\iv}X_{\nv}^\dag)
\right].
\]
After a substitution of (\ref{unitaryxy}), 
one obtain the finite difference term 
of $\eta$ 
which is a part of the fermionic kinetic term 
as follows
\[
-\frac{1}{2g^2}\sum_\nv
\Tr\left[
\frac{1}{a}\lambda_\nv(\eta_\nv-\eta_{\nv+\iv})
\right]+\cdots.
\]
One can also check the generation of other kinetic terms in the same way. 

The deconstruction does not 
change any symmetry since it is only the changing of the matrix variables.
%All symmetry
Also the BRST symmetry is unchanged 
by the deconstruction, 
its explicit form after the deconstruction is obtained 
through a rewriting of the transformation law (\ref{orb-BRST})
by using (\ref{unitaryxy}).
Indeed this is nothing but the change of the variables, but 
the difference term of $\Phi$ is realized there, for example,
\bea
Q^2 X_\nv = Q \lambda_\nv &=& 
\Phi_\nv X_\nv - X_\nv\Phi_{\nv+\iv},\nn\\
&\to& \frac{1}{a}(\Phi_\nv - \Phi_{\nv+\iv})+\cdots.\nn
\eea
Especially applying the BRST transformation twice to the
complex bosonic fields $-\frac{i}{2}(X_\nv-X^\dag_\nv)$ and $-\frac{i}{2}(Y_\nv-Y^\dag_\nv)$ 
which become the gauge fields 
$v_x$ and $v_y$
in the continuum limit,
it
generates a lattice analog of the covariant derivative $D_\mu \Phi = \partial_\mu \Phi 
+i[v_\mu, \Phi]$ ($\mu= x,y$).
This result agrees with the twice BRST transformation
of the gauge field of the continuum topological field theory,
which is represented by
\[
Q^2 v_\mu = iD_\mu \Phi.
\]

In the deconstructed action, 2 dimensional gauge coupling $g_2$
is written in terms of a product of the lattice spacing $a$
and gauge coupling of matrix model $g$ as $g_2 = g a$.
Using this 2 dimensional coupling, the sum of $N^2$ indices $\nv$ is factorized  as
\[
\frac{1}{g^2} \sum_\nv = \frac{1}{g_2^2} \left(a^2 \sum_\nv\right).
\]
Then we can regard 
$a^2\sum_\nv$ as a coordinate summation over the lattice point
with the volume element $a^2$, namely which becomes an integral 
over the 2 dimensional continuum space-time
$\int d^2 x$ with a suitable measure.

Therefore finally we can obtain the 2 dimensional topological field theory action, which is
equivalent to the $\mathcal{N} = (4,4)$ 
supersymmetric Yang-Mills theory in a sense of matter contents and interactions.
Here we should take the limit with the condition that 2 dimensional coupling 
$g_2$
is fixed.
The explicit form of the target continuum action is
%%%%% TFT form%%%%%%%%%%%%%%%%%%%%%%%%
\be
S=\frac{1}{g_2^2}\displaystyle \int d^2x 
Q\Xi(\vec{\cal B},\vec{\cal F},\Phi),
\label{TFT-cont}
\ee
where
\be
\begin{split}
\Xi(\vec{\cal B},\vec{\cal F},\Phi)
&= \Tr\Bigg[\frac{1}{4}\eta[\Phi,\Phib]
+\vec{\chi}\cdot
(\vec{H}-i\vec{\cal E})
\\
&\qquad+\frac{1}{2}
\biggl\{
(\lambda^\dag-\lambda)
D_x \Phib 
+(\lambdat^\dag-\lambdat)
D_y \Phib \\
&\qquad
+(\lambda + \lambda^\dag) [ s_0, \Phib] 
+(\lambdat + \lambdat^\dag) [ s_3, \Phib] 
\biggr\}\Bigg],
\end{split}
%\label{orb-act}
\nn
\ee
and
\bea
{\cal E}^\R &=& -2(D_x s_0 + D_y s_3),\nn\\
{\cal E}^\C &=& 2i(D_x s_3 - D_y s_0 + [s_0,s_3] + i F_{xy}),\nn
\\
F_{xy} &=& -i[D_x,D_y].\nn
\eea
This topological field theory action (\ref{TFT-cont}) is equivalent to 
${\cal N} = (4,4)$ super Yang-Mills theory action,
which is
\bea
&&S =S_B+S_F, \nn \\
&&S_B =
\frac{1}{g_2^2}
\int d^2 x\,
%\frac{1}{2} \biggl( 
%(D_{x} s_{x})^2 +
%(D_{x} s_{y})^2 +
%(D_{y} s_{x})^2 +
%(D_{y} s_{y})^2 
\Tr \left[
( D_{\mu} s_{k} )^2
+\frac{1}{4}F_{\mu\nu}F^{\mu\nu} -\frac{1}{4}\sum_{k\neq k'}[s_{k}, s_{k'}]^2
\right],\nn
%\frac{1}{2}|[s,s^\dag]|^2
%\nonumber\\
%&&+D_{x}\Phi D_{x}\bar{\Phi} 
%+D_{y}\Phi D_{y}\bar{\Phi} 
%+|[s_x, \Phi]|^2 
%+|[s_y, \Phi]|^2 
%+[\Phi,\bar{\Phi}]^2\biggr)
\\
&&S_F = 
\frac{1}{2g_2^2}
\int d^2 x\,
\Tr \biggl[
\bar{\Psi}_i \gamma^\mu D_\mu \Psi_i
+\bar{\Psi}_i [s_0, \Psi_i]
+i\bar{\Psi}_i \gamma_3 \tau^a_{ij}[s_a \Psi_j]
\biggl],\nn
%\\
%&&\text{(I have to confirm fermion value one more)} 
\eea
%where we denote $D_{m} (m =x,y)$ as covariant derivative 
%with adjoint representation 
%defined as
%$D_{m} w = \partial_{m} w + i[ v_{m}, w]$
%and coupling $g_2 = a g$ is 
%fixed during taking the continuum limit.
where indices $k, k'$ run from $0$ to $3$.
%and $F_{\mu\nu} = -i[D_\mu,D_\nu]$. 
%\\
%\section{Fermionic part}
%
%
%
Here $\gamma$-matrices are defined by
\[
\gamma_1 = -\sigma_3, \quad \gamma_2 = -\sigma_1, 
\quad \gamma_3 = \sigma_2.
\]
The fermions $\Psi_i,\bar{\Psi}_i (i =1,2)$ are given by
\[
\bar{\Psi}_1 = \left( -\chi^\C, (\frac{1}{2} \eta + i\chi^\R) \right),
\quad
\bar{\Psi}_2 = \left( \lambda, \lambdat \right), 
\quad
\Psi_1 = 
\begin{pmatrix}
\lambdat^\dag \\
-\lambda^\dag 
\end{pmatrix}
\quad
\Psi_2 = 
\begin{pmatrix}
-(\frac{1}{2} \eta - i\chi^\R) \\
-\chi^{\C\dag} 
\end{pmatrix}.
\]
And also $\tau^a_{ij}$ $(a =1,2,3)$ are Pauli matrices with flavor indices $i,j$,
and the sum of the repeated indices are taken here.
%\underline{Memo:We have to confirm whether fermion part becomes this value
%by explicit calculation}
%
%
%

\subsection{Truncation to ${\cal N}=(2,2)$ theory}

%\subsection{Hermite restriction}

So far we have been considering the theory with 8 supercharges, 
but we can obtain a subclass of lower supersymmetric
theory via a truncation of the fields where some components vanish. 
Explicitly during the deconstruction by the expansion (\ref{unitaryxy}),
%t the time when we perform deconstruction using 
we
get a topological field theory twisting from $\mathcal{N} = (2,2) $ supersymmetric Yang-Mills theory which has
4 supercharges
if we 
%the stronger constraint on the 
%$X_\nv ,X^\dag_\nv$ and $Y_\nv ,Y^\dag_\nv$
truncate some scalar and auxiliary fields
as $\chi^\R = H^\R = {\cal E}^\R = s_0=s_3=0$.
After these truncations, 
the expansion of the bosonic fields becomes
\be
\begin{array}{ll}
X_\nv =
\frac{1}{a} U_{x,\nv}, &
X^\dag_\nv =
\frac{1}{a} U_{x,\nv}^\dag,\\
Y_\nv =
\frac{1}{a} U_{y,\nv}, &
Y^\dag_\nv =
\frac{1}{a} U_{y,\nv}^\dag.
\end{array}
\label{truncation}
\ee
We here can expect that the BRST transformation of the lattice spacing $a$ vanish
since the lattice spacing is not dynamical. So from (\ref{truncation}) we immediately obtain 
\be
\begin{split}
&Q X_\nv = \frac{1}{a} Q U_{x,\nv} = \lambda_\nv,\qquad
Q (X_\nv X^\dag_\nv) = Q \frac{1}{a^2} = 0, \\
&Q Y_\nv = \frac{1}{a} Q U_{y,\nv} = \lambdat_\nv,\qquad
Q (Y_\nv Y^\dag_\nv) = Q \frac{1}{a^2} = 0.
\label{constraints for XX}
\end{split}
\ee
Using the definition (\ref{truncation}) and the BRST transformations (\ref{constraints for XX}),
we obtain the constraints between $\lambda_\nv$ ($\lambdat_\nv$) and $\lambda_\nv^\dag$ ($\lambdat_\nv^\dag$)
\be
\begin{split}
&\lambda^\dag_\nv = - U^\dag_{x,\nv} \lambda_\nv U^\dag_{x, \nv}, \\
&\lambdat^\dag_\nv = - U^\dag_{y,\nv} \lambdat_\nv U^\dag_{y,\nv}.
\end{split} \label{lambda-condition}
\ee
There also exist relationships between $\chi^\C,H^\C$ and $\chi^{\C\dag} ,
H^{\C\dag}$
which are imposed by the equations of motion $H^\C = \frac{i}{2}{\cal E}^\C$
and $H^{\C\dag} = -\frac{i}{2}{\cal E}^{\C\dag}$ and the BRST
transformations $Q\chi^\C = H^\C$ and $Q\chi^{\C\dag} = H^{\C\dag}$.
These conditions do not contradict the definition of the original
BRST transformations (\ref{orb-BRST}).
Thus we have precisely a half of the degree of freedom in the truncated theory.

Together with the above definitions and constraints, 
the truncated lattice action becomes as follows,
\be
S=\frac{1}{2g_2^2}a^2\sum_\nv Q\Xi'(\vec{\cal B}'_\nv,\vec{\cal F}'_\nv,\Phi_\nv),
%\label{quiver action 2}
\label{trun-act}
\ee
%\begin{split}
where
\[
\begin{split}
\Xi'(\vec{\cal B}'_\nv,\vec{\cal F}'_\nv,\Phi_\nv)
&= \Tr\Bigg[\frac{1}{4}\eta_{ {\bf n}}[\Phi_{ {\bf n}},\Phib_{ {\bf n}}]
+\frac{1}{2}\left( \chi^{\C \dag}_{\nv}
({H}^\C_{\nv}-i {\cal E}^\C_{\nv})
+\chi^{\C}_\nv
({H}^{\C\dag}_\nv+i{\cal E}^{\C\dag}_\nv) \right)
\\
&\qquad+\frac{1}{a}
\biggl\{
\lambda_\nv(U_{x,\nv}^\dag\Phib_\nv-\Phib_{\nv+\iv}U_{x,\nv}^\dag)
+\lambdat_\nv(U_{y,\nv}^\dag\Phib_\nv-\Phib_{\nv+\jv}U_{y,\nv}^\dag)
\biggr\}\Bigg].
\end{split} 
\]
Here indeed $\chi^{\C\dag}, H^{\C\dag}$ depend on other fields,
we use the same symbol as the ${\cal N} =(4,4)$ in this action since 
the explicit expression of these fields are too
complicated
to write down.

The truncation conditions 
%among the fermion fields  
in the continuum limit can be read from 
the $a \to 0$ limit in each truncate condition on the lattice. 
For fermion fields $\lambda,\lambdat$ and
$\lambda^\dag, \lambdat^\dag$,
the following conditions are immediately obtained from 
(\ref{lambda-condition})
%can be read off as
\[
\lambda = -\lambda^\dag, \qquad
\lambdat = -\lambdat^\dag.
\]
%\\
We also get a 
constraint between
$H^\C$ and $H^{\C\dag}$ as
\be
H^\C = -H^{\C\dag} \label{H-cond}
\ee
since 
$H^\C$ is given as $\frac{i}{2}{\cal E}^\C$
which tends to be the field strength $-iF_{xy}$ in the continuum limit
while $H^\C$ is defined by $-\frac{i}{2}{\cal E}^{\C\dag} \sim iF_{xy}$.
%the constraint between the $\chi^\C$ and $\chi^{\C\dag}$
Using this condition we can obtain the constraint 
for their BRST partners $\chi^\C$ and  
$\chi^{\C\dag}$.
From 
the above condition (\ref{H-cond}) and
$a \to 0$ limit of
(\ref{orb-BRST}),
we can see following conditions 
\bea
%\begin{array}{ll}
Q (H^\C + H^{\C\dag}) &\to& [\Phi, \chi^\C +\chi^{\C\dag}] =0, \label{commu-chi}
\\
Q (\chi^\C +\chi^{\C\dag}) &\to& H^\C + H^{\C\dag}  = 0.\label{fermi-close}
%\end{array}
\eea
The commutation relation (\ref{commu-chi}) tells us that
$\chi^\C+ \chi^{\C\dag}$ must be proportional to the fermionic unit 
matrix which decouples from the continuum 
action. 
% written by the adjoint fields.
%Here $chi^\C + \chi^{\C\dag}$ cannot be represented 
%by the other field variable because we cannot
%the dynamical 
So we can impose %ignore the $\chi^\C +\chi^{\C\dag}$ at continuum limit,
\be
\chi^\C = -\chi^{\C\dag},
\ee
at the action construction.
Moreover we can obtain the gauge fields from
the $-i (X-X^\dag)$ and
$-i (Y-Y^\dag)$ which are the same
ones in ${\cal N} =(4,4)$ theory.
%In the continuum limit, these apparently becomes:
Combining these conditions, 
we find the BRST transformations 
%in the continuum theory
\[
\begin{array}{ll}
Qv_{\mu}=\lambda_{\mu}, & Q\lambda_{\mu} =iD_\mu\Phi,\\
QH = [\Phi,\chi], & Q\chi = H,\\
Q\Phib=\eta, & Q\eta=[\Phi,\Phib],\\
Q\Phi=0, & 
\end{array}
\]
where 
\[
\begin{split}
\lambda_x &= -\frac{i}{2} ( \lambda -\lambda^\dag), \\
\lambda_y &= -\frac{i}{2} ( \lambdat -\lambdat^\dag), \\
\chi &= -\frac{i}{2}(\chi^\C -\chi^{\C\dag}), \\
H &= -\frac{i}{2}(H^\C -H^{\C\dag}). 
\end{split}
\]
So the continuum limit of the truncated action (\ref{trun-act})
becomes a topological field theory action
twisted from the ${\cal N} = (2,2)$ 
2 dimensional supersymmetric Yang-Mills 
theory.
%Due to those,
The action is
\be
S=\frac{1}{2g_2^2}\int d^2 x \, Q\Xi'(\vec{\cal B}',\vec{\cal F}',\Phi),
\label{continuum action}
\ee
with
\[
\Xi'(\vec{\cal B}',\vec{\cal F}',\Phi)
= \Tr\Bigg[\frac{1}{4}\eta[\Phi,\Phib]
+\chi
(H-i{\cal E})
-i%\sum_{\mu=x,y}
\lambda^{\mu} D_\mu\Phib
\Bigg],
\]
where
\[
{\cal E} = -2\left\{
\del_x v_{y} -\del_y v_{x}
+i[v_{x},v_{y}]
\right\}=-2F_{xy}.
\]
%This action is nothing but 

Our lattice theory is not equivalent to 
the
${\cal N} =(2,2)$ lattice theory proposed by 
Cohen, Kaplan, Katz and Unsal (CKKU) \cite{Cohen:2003xe}
although the continuum limits of the both coincide.
Our model has only one BRST closed field
$\Phi$ while the CKKU model has two BRST closed fields
$\bar{x}$ and $\bar{y}$.
Each fermionic charge in the models cannot be equivalent 
to each other.
In addition, our model does not have 3-point scalar vertices,
which are included in the CKKU model.

%\subsection{Unitary restriction}
%\subsection{fermion determinant}

%\section{Decent relation}
%%%%%%%%%%%%%%%%%%%%%%%%%%%%%%%%%%%%%%%%%%%%%%%%%%%%%%%%%%%%%
%                                                           %
%   We have to discuss about this section                   %
%                                                           %
%%%%%%%%%%%%%%%%%%%%%%%%%%%%%%%%%%%%%%%%%%%%%%%%%%%%%%%%%%%%%
\section{Partition Function and Observables}\label{Sec_PO}
\subsection{Partition function}\label{Sec:partition}
As we have seen, the action of the reduced matrix model can be written in the BRST exact form.
So if we consider the partition function of the theory
\[
Z=\frac{1}{{\rm Vol}(U(MN^2))}
\int_{\cal M} [d\vec{\cal B}] [d\vec{\cal F}] [d\Phi]\, e^{-\frac{1}{2g^2}Q\Xi(\vec{\cal B},\vec{\cal F},\Phi)},
\]
where the path integral is performed with a suitable measure on the moduli space ${\cal M}$,
and ${\rm Vol}(U(MN^2))$ represents a volume of $U(MN^2)$ group (see for example
the explicit form and asymptotic behavior in 
\citen{Ooguri:2002gx}).
Noting that a derivative of the partition function with respect to the gauge coupling $g$ vanishes
%\be
%\begin{split}
%\frac{\del Z}{\del g}&=g^{-3}\int_{\cal M} [d\vec{\cal B}] [d\vec{\cal F}] \,
%(Q \Xi) e^{-\frac{1}{2g^2}Q\Xi}\\
%&=g^{-3}\int_{\cal M} [d\vec{\cal B}] [d\vec{\cal F}] \,
%Q(\Xi e^{-\frac{1}{2g^2}Q\Xi})=0,
%\end{split}
%\ee
\[
\begin{split}
\frac{\del Z}{\del g}&\propto \int_{\cal M} [d\vec{\cal B}] [d\vec{\cal F}] 
[d\Phi]\,
(Q \Xi) e^{-\frac{1}{2g^2}Q\Xi}\\
&=\int_{\cal M} [d\vec{\cal B}] [d\vec{\cal F}] [d\Phi]\,
Q(\Xi e^{-\frac{1}{2g^2}Q\Xi})=0,
\end{split}
\]
since
%$\langle Q \Xi
%\exp\{-\frac{1}{2g^2}Q\Xi\}\rangle =0$ and
the measure is defined as the BRST invariant,
the partition function itself does not depend on the coupling $g$.
If we evaluate the partition function in the limit of 
%$g\rightarrow \infty$,
$g\rightarrow 0$,
the semi-classical (Gaussian, WKB) approximation is exact.
This is a property of topological field theory. 

\subsubsection{Exact partition function\\}

The calculation technique was introduced by 
\citen{Moore:1997dj,Moore:1998et}
and developed in
\citen{Bruzzo:2002xf,Fucito:2004gi}.
We use the supersymmetric version of the localization theorem and
most arguments are followed by \citen{Bruzzo:2002xf,Fucito:2004gi}.
One would see that the technique is analogous
to the Nicolai mapping \cite{Sakai:1983dg}.

We first introduce a vector field $Q^*$ generating the BRST transformations
on the supermanifold  spanned by the coordinates $\vec{\cal A}=(\vec{\cal B},\vec{\cal F})$
\be
\begin{split}
Q^* &=
\lambda\frac{\del}{\del X} + \lambda^\dag\frac{\del}{\del X^\dag}
+\lambdat\frac{\del}{\del Y}+\lambdat^\dag\frac{\del}{\del Y^\dag}
+[\Phi,\vec{\chi}]\cdot\frac{\del}{\del \vec{H}}
+\eta\frac{\del}{\del \Phib}\\
&\quad
+[\Phi,X]\frac{\del}{\del \lambda}
+[\Phi,X^\dag]\frac{\del}{\del \lambda^\dag}
+[\Phi,Y]\frac{\del}{\del \lambdat}
+[\Phi,Y^\dag]\frac{\del}{\del \lambdat^\dag}\\
&\qquad
+\vec{H}\cdot \frac{\del}{\del \vec{\chi}}+[\Phi,\Phib]\frac{\del}{\del \eta}\\
&\equiv \vec{Q}^*_{\cal F}\cdot \frac{\del}{\del \vec{\cal B}}
+\vec{Q}^*_{\cal B} \cdot \frac{\del}{\del \vec{\cal F}}.
\end{split}
\label{BRST vector field}
\ee
The super-Hessian associated with the vector field $Q^*$ is given by
\[
{\cal L}(\Phi)=\left(
\begin{array}{cc}
\frac{\del (Q^*_{\cal B})^i}{\del {\cal B}^j} & \frac{\del (Q^*_{\cal B})^i}{\del {\cal F}^j}\\
\frac{\del (Q^*_{\cal F})^i}{\del {\cal B}^j} & \frac{\del (Q^*_{\cal F})^i}{\del {\cal F}^j}
\end{array}
\right).
\]
According to the localization theorem \cite{Bruzzo:2002xf,Fucito:2004gi},
after performing the integrals on ${\cal M}$ except for $\Phi$,
the partition function is given in terms of a super-determinant of the super-Hessian
\[
\begin{split}
Z&=\frac{1}{{\rm Vol}(U(MN^2))}\int [d\vec{\cal B}] [d\vec{\cal F}] [d\Phi]
e^{-\frac{1}{g^2}Q\Xi(\vec{\cal B},\vec{\cal F},\Phi)}\\
&=\frac{1}{{\rm Vol}(U(MN^2))}\int [d\Phi] \frac{1}{{\rm Sdet}^{1/2}\,{\cal L}(\Phi)}.
\end{split}
\]

However if we naively apply the above derivation to the present case,
the integral measure on $\Phi$ divided by the super-determinant is degenerate
and the integral diverges.
In order to regularize the integral, the authors of \citen{Moore:1997dj,Moore:1998et}
introduce parameters $\e$ and $\et$ ($\Omega$-background)
 which give masses for the matrices %$(\vec{\cal B},\vec{\cal F})$
and lift up the flat directions. The BRST transformations are modified by
the parameters as
\be
\begin{array}{ll}
Q_\e X=\lambda, & Q_\e \lambda =[\Phi,X]+\e X,\\
Q_\e Y=\lambdat, & Q_\e \lambdat =[\Phi,Y]+\et Y,\\
Q_\e \vec{H}=[\Phi,\vec{\chi}]+\vec{\e}\cdot\vec{\chi}, & Q_\e \vec{\chi} = \vec{H},\\
Q_\e \Phib=\eta, & Q_\e \eta=[\Phi,\Phib],\\
Q_\e \Phi=0, & \label{BRST-def_e}
\end{array} 
\ee
where $\vec{\e}\equiv(0,\e+\et,-(\e+\et))$.
Following these modified BRST transformations,
the vector fields are also modified as
\[
\begin{split}
Q_\e^* &=
\lambda\frac{\del}{\del X} + \lambda^\dag\frac{\del}{\del X^\dag}
+\lambdat\frac{\del}{\del Y}+\lambdat^\dag\frac{\del}{\del Y^\dag}
+([\Phi,\vec{\chi}]+\vec{\e}\cdot \vec{\chi})\cdot\frac{\del}{\del \vec{H}}
+\eta\frac{\del}{\del \Phib}\\
&\quad
+([\Phi,X]+\e X)\frac{\del}{\del \lambda}
+([\Phi,X^\dag]-\e X^\dag)\frac{\del}{\del \lambda^\dag}
+([\Phi,Y]+\et Y)\frac{\del}{\del \lambdat}
+([\Phi,Y^\dag]-\et Y^\dag)\frac{\del}{\del \lambdat^\dag}\\
&\qquad
+\vec{H}\cdot \frac{\del}{\del \vec{\chi}}+[\Phi,\Phib]\frac{\del}{\del \eta}\\
&\equiv \vec{Q}^*_{\e{\cal F}}\cdot \frac{\del}{\del \vec{\cal B}}
+\vec{Q}^*_{\e{\cal B}} \cdot \frac{\del}{\del \vec{\cal F}},
\end{split}
\]
and correspondingly
the super-determinant is expressed explicitly as
\[
\Sdet {\cal L}_\e(\Phi)
=\frac{\det([\Phi,\cdot]^2-\e^2)
\det([\Phi,\cdot]^2-\et^2)}
{\det([\Phi,\cdot]^2-(\e+\et)^2)},
\]
where $[\Phi,\cdot]$ stands for an adjoint action causing by $\Phi$.
If we diagonalize $\Phi$ as
\[
\Phi \rightarrow \diag(\phi_1,\phi_2,\ldots,\phi_N),
\]
the adjoint action and a part of the integral measure are replaced by the so-called
Vandermonde determinant.
Then we finally obtain the partition function as an integral over the eigenvalues of $\Phi$
\be
Z=\frac{(\e+\et)^{MN^2}}{\e^{MN^2}\et^{MN^2}}\int \prod_{i=1}^{MN^2} d\phi_i
\prod_{1 \leq i<j \leq MN^2}
\frac{(\phi_i-\phi_j)^2((\phi_i-\phi_j)^2-(\e+\et)^2)}
{((\phi_i-\phi_j)^2-\e^2)((\phi_i-\phi_j)^2-\et^2)}.
\ee

Now let us apply the above formulation to our orbifold model given by
the BRST transformations (\ref{orb-BRST}) and the action (\ref{quiver action}).
From the BRST transformations, we find the vector field $Q^*$ as
\be
\begin{split}
Q^* &=\sum_\nv\Bigg[
\lambda_\nv\frac{\del}{\del X_\nv} + \lambda_\nv^\dag\frac{\del}{\del X_\nv^\dag}
+\lambdat_\nv\frac{\del}{\del Y_\nv}+\lambdat_\nv^\dag\frac{\del}{\del Y_\nv^\dag}\\
&\qquad\quad
+[\Phi_\nv,\chi^\R_\nv]\frac{\del}{\del H^\R_\nv}
+(\Phi_\nv\chi^\C_\nv-\chi^\C_\nv\Phi_{\nv+\iv+\jv})\frac{\del}{\del H^\C_\nv}
+(\Phi_\nv\chi^{\C\dag}_\nv-\chi^{\C\dag}_\nv\Phi_{\nv-\iv-\jv})
\frac{\del}{\del H^{\C\dag}_\nv}
+\eta_\nv\frac{\del}{\del \Phib_\nv}\\
&\qquad\quad
+(\Phi_\nv X_\nv - X_\nv \Phi_{\nv+\iv})\frac{\del}{\del \lambda_\nv}
+(\Phi_\nv X_\nv^\dag - X_\nv^\dag \Phi_{\nv-\iv})\frac{\del}{\del \lambda_\nv^\dag}\\
&\qquad\quad
+(\Phi_\nv Y_\nv - Y_\nv \Phi_{\nv+\jv})\frac{\del}{\del \lambdat_\nv}
+(\Phi_\nv Y_\nv^\dag - Y_\nv^\dag \Phi_{\nv-\jv})\frac{\del}{\del \lambdat_\nv^\dag}\\
&\qquad\quad
+\vec{H}_\nv\cdot \frac{\del}{\del \vec{\chi}_\nv}
+[\Phi_\nv,\Phib_\nv]\frac{\del}{\del \eta_\nv}
\Bigg]\\
&\equiv \sum_\nv\left[
\vec{Q}^*_{{\cal F}_\nv}\cdot \frac{\del}{\del \vec{\cal B}_\nv}
+\vec{Q}^*_{{\cal B}_\nv} \cdot \frac{\del}{\del \vec{\cal F}_\nv}
\right].
\end{split}
\ee
So we can express the partition function in terms of integrals on
the $M$ eigenvalues $\phi_{\nv,i}$ of each $\Phi_\nv$
\be
\begin{split}
Z&=\frac{1}{{\rm Vol}(U(M))^{N^2}}\int \prod_\nv
[d\vec{\cal B}_\nv] [d\vec{\cal F}_\nv] [d\Phi_\nv]
e^{-\frac{1}{g^2}\sum_\nv Q\Xi(\vec{\cal B}_\nv,\vec{\cal F}_\nv,\Phi_\nv)}\\
&=\frac{1}{{\rm Vol}(U(M))^{N^2}}\int \prod_\nv [d\Phi_\nv]
\frac{1}{{\rm Sdet}^{1/2}\,{\cal L}(\Phi_\nv)}\\
&=\int \prod_\nv \left[
\prod_{i=1}^M d\phi_{\nv,i}
\prod_{1 \leq i \leq j \leq M}
\frac{(\phi_{\nv,i}-\phi_{\nv,j})^2(\phi_{\nv,i}-\phi_{\nv+\iv+\jv,j})^2}
{(\phi_{\nv,i}-\phi_{\nv+\iv,j})^2(\phi_{\nv,i}-\phi_{\nv+\jv,j})^2}
\right]. \label{partition-func-lat}
\end{split}
\ee
Notice that we do not need the parameters $\e$ and $\et$ no longer
in order to regularize the partition function. The orbifolding plays role
instead of the $\Omega$-background.

%One observable
%\be
%\left\langle
%e^{-\frac{\mu}{2}\sum_\nv \Tr \Phi_\nv}
%\right\rangle
%=\frac{1}{Z}
%\int \prod_\nv \left[
%\prod_{i=1}^M d\phi_{\nv,i}
%\prod_{1 \leq i \leq j \leq M}
%\frac{(\phi_{\nv,i}-\phi_{\nv,j})^2(\phi_{\nv,i}-\phi_{\nv+\iv+\jv,j})^2}
%{(\phi_{\nv,i}-\phi_{\nv+\iv,j})^2(\phi_{\nv,i}-\phi_{\nv+\jv,j})^2}
%e^{-\frac{\mu}{2}\sum_{i=1}^M \phi_{\nv,i}}
%\right]
%\ee

%\paragraph{The case of ${\cal N} = (2,2)$ theory\\}
%
%Here we will obtain the partition function of ${\cal N} = (2,2)$
%supersymmetric Yang-Mills theory.
%To apply our technique for ${\cal N} = (2,2)$ case, 
%we should consider the measure condition with
%$\lambda = -\lambda^{\dag}$, $\lambdat = -\lambdat^{\dag}$, 
%$\chi^\C = -\chi^{\C \dag}$, $\chi^\R =0$.

\subsection{Observable}
A physical observable ${\cal O}$ 
in the topological field theory is defined as a BRST
closed operator 
\[
Q{\cal O}=0,
\]
but not BRST exact. Then the observables form the BRST cohomologies.
The expectation value of the BRST operator is also independent 
of the coupling constant $g$
using the same arguments on the partition function.
So we can use the semi-classical approximation to calculate
the expectation value of the physical operators.

Corresponding observables on our lattice theory would be exactly defined
as the BRST cohomology
on the lattice 
%formulation 
since the action of this theory can be written as BRST exact. 
%which are same as one 
%on the continuum topological field theory.
In our construction, 
the BRST cohomology on the lattice theory
comes from the BRST cohomology in the matrix model.
So in this paper, we look for the BRST cohomological value 
in the matrix model
to define the observables on the lattice. 
%at first.
%And 
%from that 
%We also discuss a role of \textit{deconstruction}
%on defining the BRST cohomology.

\subsubsection{The BRST cohomology on the continuum space-time\\}

To investigate the BRST cohomology in the matrix model,
we first would like to mention about 
the BRST cohomology in continuum theory \cite{Witten:1988ze}.
Here we pick up the 4 dimensional ${\cal N} =2$
topological field theory since 
our matrix model is obtained by the dimensional reduction 
of it as denoted in section 2. 
In continuum 
theory, we can obviously find a candidate for a gauge invariant
physical observable
\[
{\cal O}_0 =\frac{1}{2}
%\displaystyle \int_{\Sigma} d\mu
\Tr \Phi^2(x),
\]
or a trace of a polynomial in $\Phi(x)$ generally, since the field $\Phi$ is BRST closed but not BRST exact. 
%Here we denote $\int_{\Sigma} d\mu$ as the integration over the differential 
%manifold 
%The operator ${\cal O}_0$ is a scalar function and associated with a 0-form.
%We also can have higher form BRST closed operators ${\cal O}_1,{\cal O}_2,\cdots$.
For higher form operators, noting the following relation
\[
(d+Q)\frac{1}{2}\Tr (F+\lambda+\Phi)^2 =0,
\]
where 
$F=F_{\mu\nu}dx^\mu\wedge dx^\nu$ is Yang-Mills field strength,
$\lambda=\lambda_\mu dx^\mu$ is a fermionic 1-form and 
$d$ is an external derivative,
we obtain the so-called descent equations
\be
\begin{split}
Q{\cal W}_0&=0,\\
d{\cal W}_0 + Q{\cal W}_1&=0,\\
d{\cal W}_1 + Q{\cal W}_2&=0,\\
d{\cal W}_2 + Q{\cal W}_3&=0,\\
d{\cal W}_4 &= 0,
\end{split}
\label{descent}
\ee
%for 4 dimensional topological field theory, 
where
\[
\begin{split}
&{\cal W}_0=\frac{1}{2}\Tr \Phi^2\\
&{\cal W}_1=\Tr \Phi\lambda\\
&{\cal W}_2=\Tr (\Phi F+\frac{1}{2}\lambda\wedge\lambda)\\
&{\cal W}_3=\Tr F\lambda\\
&{\cal W}_4=\frac{1}{2}\Tr F\wedge F.
\end{split}
\]
Using a property of the descent equations (\ref{descent}),
the observables can be constructed from the $k$-form ${\cal W}_k$ 
$(k =1, \ldots, 4)$
by picking up a $k$-homology cycle $\gamma_k$ as follows
\[
{\cal O}_k = \int_{\gamma_k} {\cal W}_k,
\]
since
\[
Q{\cal O}_k=-\int_{\gamma_k} d{\cal W}_{k-1}=-\int_{\del\gamma_k} 
{\cal W}_{k-1}=0.
\]
So ${\cal O}_k$ can be the BRST closed observables.
% which is BRST closed. 
%but cannot be written as
%a BRST exact form of a gauge invariant 
%operator. 
%\\
%\underline{Memo: Should we denote following sentences?}\\
%Here 
%we note 
%these operators are the cohomology of 
%are 
%moduli space ${\cal M_F}$ 
%which are the space of flat connections
%such that
%\be
%${\cal O}_0 \in H^4({\cal M_F})$,${\cal O}_1 \in H^3({\cal M_F})$,
%${\cal O}_2 \in H^2({\cal M_F})$.
%And we can regard $Q$ as the exterior derivative on the ${\cal M_F}$.

\subsubsection{The BRST cohomology in the matrix model\\}
%Our purpose in this paper is 
%to construct the lattice theory of 
%two dimensional topological field
%theory by the \textit{'deconstruction'}.
%
%however, 
There is expectation that the BRST
closed observables are obtained similarly by using
the descent equations in the matrix model.
%since the reduced matrix model is 0 dimensional theory.
From a dimensional reduction of 
(\ref{descent}), we obtain 
\be
\begin{array}{ll}
{\cal W}_{1,\mu} = \Tr\Phi\lambda_\mu, &
\qquad{\cal W}_{2,\mu\nu}= \Tr\left( \Phi[A_\mu,A_\nu] 
- \lambda_\mu\lambda_\nu \right),
\end{array}\label{original}
\ee
where the forms $dx^\mu$ or $dx^\mu \wedge dx^\nu$ does not
make a sense
in 0 dimension and $\mu$, $\nu$ run from 1 to 4.
It is more convenient to redefine the above 
values as the eigenstates of the
$r_a$ charges denoted in (\ref{r-charge-def}) for our purpose.
Combining together ${\cal W}_{1,\mu}$, 
we define the following values
\be
\begin{array}{ll}
{\cal W}^x_1 = \Tr\Phi\lambda,&
\qquad{\cal W}^{\bar{x}}_1 = \Tr\Phi\lambda^\dag,\\
{\cal W}^y_1 = \Tr\Phi\lambdat,&
\qquad{\cal W}^{\bar{y}}_1 = \Tr\Phi\lambdat^\dag.
\end{array}\label{1-f-om}
\ee
From the combination of ${\cal W}_{2,\mu\nu}$, 
we also define the following values 
\be
\begin{array}{ll}
{\cal W}^{x\bar{x}}_2=
\Tr(\Phi[X,X^\dag]
-\lambda\lambda^\dag),&
\quad{\cal W}^{y\bar{y}}_2=
\Tr(\Phi[Y,Y^\dag]
-\lambdat\lambdat^\dag),\\
{\cal W}^{xy}_2=
\Tr(\Phi[X,Y]
-\lambda\lambdat),&
\quad{\cal W}^{\bar{x}\bar{y}}_2=
\Tr(\Phi[X^\dag,Y^\dag]
-\lambda^\dag\lambdat^\dag).
\end{array}\label{2-f-om}
\ee

The dimensional reduction of the descent equation 
(\ref{descent}) tells us that the values 
(\ref{original}) become BRST closed, namely
$Q{\cal W}_{1,\mu} = 0$ and $Q{\cal W}_{2, \mu\nu} = 0$,
since there is no concept of the differential form
and the derivative in the 0 dimensional matrix model.
So these (\ref{original}) and 
their combinations (\ref{1-f-om}) and (\ref{2-f-om}) are manifestly
BRST closed.

But unfortunately, such the matrix operators 
(\ref{original}), (\ref{1-f-om}) and (\ref{2-f-om})
associated to higher forms
can also be written in the BRST exact form as
\[
\begin{array}{ll}
{\cal W}_{1,\mu} = Q\Tr\Phi A_\mu, &
\qquad{\cal W}_{2,\mu\nu}= Q\Tr \lambda_\mu A_\nu,
\end{array}%\label{original},
\]
\[
\begin{array}{ll}
{\cal W}^x_1 = Q\Tr\Phi X,&
\qquad{\cal W}^{\bar{x}}_1 = Q\Tr\Phi X^\dag,\\
{\cal W}^y_1 = Q\Tr \Phi Y,&
\qquad{\cal W}^{\bar{y}}_1 = Q\Tr\Phi Y^\dag,
\end{array}
%\label{O1 exact form}
\]
and
\[
\begin{array}{ll}
{\cal W}^{x\bar{x}}_2=Q
\Tr\lambda X^\dag,&
\quad{\cal W}^{y\bar{y}}_2=Q
\Tr\lambdat Y^\dag,\\
{\cal W}^{xy}_2=Q
\Tr\lambda Y,&
\quad{\cal W}^{\bar{x}\bar{y}}_2=Q
\Tr\lambda^\dag Y^\dag.
\end{array}
%\label{O2 exact form}
\]
Therefore these (\ref{original}), (\ref{1-f-om}) and (\ref{2-f-om}) cannot be 
observable. 
Only the trace of the polynomial of $\Phi$ 
can be 
non-trivial
observable. In particular we choose
\[
{\cal O}_0 = {\cal W}_0 = \frac{1}{2}\Tr \Phi^2,
\]
which has ghost number 4.

%Such triviality comes from 
In our model, 
all matrix operators associated with higher 
forms cannot be non-trivial observables since 
the BRST transformations in the matrix model
are  homogeneous on
 $\vec{\cal A} = (\vec{\cal B},\vec{\cal F})$.
We will show it in the following sentences.
%Since $Q$ is 
%defined as the
%homogeneous transformation about the $\vec{\cal A}$,
%%we cannot obtain the
%there is not 
%non-trivial $Q$-cohomological value 
%in the reduced matrix model
%except for 
%${\cal O}_0$ which is polynomial composed only by $\Phi\, (Q\Phi =0)$.
%We will show that when $Q$ is defined as 
%the homogeneous transformation of 
\\
\\
%From the property of the BRST transformation 
%defined as homogeneous transformation of 
%$\vec{\cal A} = (\vec{\cal B},\vec{\cal F})$,
%we can write down the BRST transformations
%by the tangent vector on $\vec{\cal A}$ as
%\bea
%Q &=&[\Phi, X]\frac{\partial}{\partial \lambda}
%+[\Phi, X^\dag]\frac{\partial}{\partial \lambda^\dag}
%+[\Phi, Y]\frac{\partial}{\partial \lambdat}
%+[\Phi, Y^\dag]\frac{\partial}{\partial \lambdat^\dag}
%+[\Phi, \bar{\Phi}]\frac{\partial}{\partial \eta}
%+[\Phi, \vec{\chi}]\cdot\frac{\partial}{\partial \vec{H}} \nonumber\\
%&& +\lambda \frac{\partial}{\partial X}
%+\lambda^\dag \frac{\partial}{\partial X^\dag}
%+\lambdat \frac{\partial}{\partial Y}
%+\lambdat^\dag \frac{\partial}{\partial Y^\dag}
%+\eta \frac{\partial}{\partial \bar{\Phi}}
%+\vec{H}\cdot \frac{\partial}{\partial \vec{\chi}}.
%\eea
In addition to the homogeneous BRST symmetry 
%which can be written by the vector denoted in 
(\ref{BRST vector field}),
we now define the another fermionic transformation $\tilde{Q}$ as
\[
\tilde{Q} = 
X \frac{\partial}{\partial \lambda}
+X^\dag \frac{\partial}{\partial \lambda^\dag}
+Y \frac{\partial}{\partial \lambdat}
+Y^\dag \frac{\partial}{\partial \lambdat^\dag}
+\bar{\Phi} \frac{\partial}{\partial \eta}
+\vec{\chi}\cdot \frac{\partial}{\partial \vec{H}}
,
\]
where this additional transformation does not have to be
symmetry of the matrix model action.
Then the anticommutation relation
between the BRST charge and additional charge $\tilde{Q}$
composes 
a number operator $\hat{N}_{\cal A}$
acting on 
$\vec{\cal A}$, which is written by
\bea
\{ Q, \tilde{Q}\}
&=&
X \frac{\partial}{\partial X}
+X^\dag \frac{\partial}{\partial X^\dag}
+Y \frac{\partial}{\partial Y}
+Y^\dag \frac{\partial}{\partial Y^\dag}
+\bar{\Phi} \frac{\partial}{\partial \bar{\Phi}}
+\vec{H}\cdot \frac{\partial}{\partial \vec{H}}
\nonumber\\
&&+\lambda \frac{\partial}{\partial \lambda}
+\lambda^\dag \frac{\partial}{\partial \lambda^\dag}
+\lambdat \frac{\partial}{\partial \lambdat}
+\lambdat^\dag \frac{\partial}{\partial \lambdat^\dag}
+\eta \frac{\partial}{\partial \eta}
+\vec{\chi}\cdot \frac{\partial}{\partial \vec{\chi}}\nonumber\\
&=&\hat{N}_{\cal A}.
\label{number}
\eea
%We note that such additional 
A general 
%$Q$-closed 
function of the matrices  $F$ can be written in terms of  
a sum of eigenfunction of $\hat{N}_{\cal A}$,
namely
%where
\be
\begin{array}{l}
F = \sum_{n_{\cal A}=0}^{\infty} F_{n_{\cal A}},\\
\hat{N}_{\cal A} F_{n_{\cal A}} = n_{\cal A}F_{n_{\cal A}},\quad
n_{\cal A} \in \N.
\end{array}
\label{expand}
\ee
%where $F_{n_{\cal A}}$ is defined as the polynomial
%The sum of 
%we can say following property for any $Q$-closed function  
%with $m$ degree of $\vec{\cal A}$,  
From (\ref{expand}) and a property 
$[Q, \hat{N}_{\cal A}]=0$,
%such 
any BRST closed function $h$ must satisfy
\be
Q h 
%= Q \left( h_0 + h_1 + \cdots \right)
= \sum_{n_{\cal A} = 0}^{\infty}  Q h_{n_{\cal A}} = 0
\Leftrightarrow Qh_{n_{\cal A}} = 0. %\quad ({}^\forall n_{\cal A} 
%\in \N ).
\label{independent}
\ee
%Due to the
% fact
%number operator which is composed by BRST charge,
%So from (\ref{number}),
%for 
The BRST closed eigenfunction $h_{n_{\cal A}}$
associated with a non zero eigenvalue $n_{\cal A} \ne 0$
must be 
BRST exact 
%value 
%in our model
since we can write such the function $h_{n_{\cal A}}$ as
%we can see
\be
h_{n_{\cal A}} 
= n_{\cal A}^{-1}
\hat{N_{\cal A}}\cdot h_{n_{\cal A}}
= n_{\cal A}^{-1}
\{ Q , \tilde{Q} \} h_{n_{\cal A}}
= n_{\cal A}^{-1}Q\cdot(\tilde{Q}h_{n_{\cal A}}).
\label{exact}
\ee
%So $Q$-closed $F_m (m\ne 0)$  
So from (\ref{independent}) and (\ref{exact}), 
BRST cohomologies
in the BRST closed function
$h = \sum_{n_{\cal A}=0}^{\infty} h_{n_{\cal A}}$ come only from the 
zero eigenstates $h_0$, which is a polynomial 
of $\Phi$ which does not associate with higher forms.
So it is shown that there is no higher form operator
which can be non-trivial observable.
%Before we close this proof,
%we comment that we can always write down the 
%BRST transformation as the tangent vector on certain functional subspace 
%$\vec{\cal R}$ if BRST charge is defined as the homogeneous transformation
%of $\vec{\cal R}$.
%And there is always an additional fermionic charge 
%which realize the number operator acting on 
%$\vec{\cal R}$ by the anticommutation relation between the BRST charge.
%(End of proof)
\\
\\
This proof tells us that
we cannot 
straightforwardly
realize the observables except for the polynomial 
of $\Phi$ on the lattice
since this situation
does not change even after performing orbifolding and 
deconstruction.
%by the higher form.
So we have to perform some non-trivial trick to realize 
the observables
%topological quantities 
on this lattice construction.

\subsubsection{Discussion about the observable}
To tell the truth, even in the continuum theory, 
the operators ${\cal O}_k(x)$
{\it formally} can be written as the BRST exact form like as 
\bea
&&{\cal O}_1 = \displaystyle \int_{\gamma_1} {\cal W}_1 
=Q \displaystyle \int_{\gamma_1}  \Tr \Phi A,
\nonumber\\
&&{\cal O}_2 = 
\displaystyle \int_{\gamma_2} {\cal W}_2 
= Q \displaystyle \int_{\gamma_2}\Tr \lambda \wedge A,
\nonumber
\eea
here $A = A_{\mu} dx^{\mu}$ is 1-form defined with the
gauge fields $A_\mu$.
These are seemingly BRST exact but actually they are not,
since
$\Tr \Phi A$ and $\Tr \lambda \wedge A$ are
not gauge invariant due to the existence of
inhomogeneous terms $g d g^{-1}$
in the gauge transformation of gauge fields $A \to g A g^{-1} 
+ g d g^{-1}$.
%(See the discussions in \cite{Kanno et al}.).
Moreover the BRST transformations in continuum theory are not
homogeneous due to such the inhomogeneous terms.
The inhomogeneous terms are absent in the matrix model
since these terms are defined with derivatives which disappear 
in 0 dimensional theory.
So in the matrix model, the higher forms are written in terms of the fields 
which are obtained by the BRST transformation of gauge invariant 
ones, they become trivial.

%\underline{sub draft}\\
In continuum theory, a vacuum expectation value of the operator
${\cal O}_0 = \Tr \Phi^2(x)$
is independent of the space-time coordinate because of the decent
relation,
\bea
 d \langle \frac{1}{2} (\Tr \Phi^2 (x) ) \rangle
&=& \langle \Tr \Phi(x) d \Phi (x) \rangle
\nonumber\\
&=& \langle \Tr \Phi(x) d_A \Phi(x) \rangle
\nonumber\\
&=& \langle Q \Tr \Phi(x) \lambda(x) \rangle
\nonumber\\
&=& 0. \nonumber
\eea
%We think 
This is
a reason why ${\cal O}_0 = {\cal W}_0 = \Tr \Phi^2$
can be defined in the reduced matrix and deconstructed lattice models.
On the other hand, 1-form and 2-form observables are defined 
as the integrations over each homology 1-cycle 
$\gamma_1 \in H_1(\Sigma )$
and 2-cycle
$\gamma_2 \in H_2(\Sigma )$
on the differentiable 2 dimensional manifold $\Sigma$,
such the integrations are the inner products $\Upsilon$
between
$\gamma_1$ or $\gamma_2$ and
their dual cohomologies $\omega_1 \in H^1(\Sigma)$ or 
$\omega_2 \in H^2(\Sigma)$,
%written as
namely
\[
\Upsilon: H_k(\Sigma) \times H^k(\Sigma) \to \R,
\qquad
\Upsilon (\gamma_k, \omega_k) = \displaystyle \int_{\gamma_k} \omega_k,
\quad (k =1,2).
\] 
The inner product between a homology cycle and its dual space
is difficult to construct in the lattice or matrix model
since such value is related to a 
topology which is invariant under
the infinitesimal transformation on the 
continuum manifold, 
such the infinitesimal transformation and its invariance 
are hard to realize in the matrix or
discrete lattice model. 
The exterior derivative, which is necessary
for the
definition of the de Rham cohomology on $\Sigma$, 
is also difficult to realize since there is 
no concept of differential form and the 
Leibnitz rule is broken on the lattice.
These would be reasons why
there are not non-trivial higher form observables
on the lattice.

\section{Conclusion and Discussions}
\label{Sec_Conc}

In this paper we have discussed about the 
partition function and observables of 
the topological field theory on the lattice.
%using by our original lattice action. 
%which can possess fermionic charge on it.
There we use our original lattice action 
%with preserved BRST charge 
which
is constructed from the 0 dimensional matrix model
%using 
using by the orbifolding and the deconstruction method, and
preserves the BRST charge on the lattice.
%Our original action is constructed 
%it 
%and this lattice can preserve the fermionic charge on it.
%with the topological twist different from \citen{Cohen:2003qw}.
The starting 0 dimensional matrix model is
the same as the ``mother theory''
introduced in \citen{Cohen:2003qw},
but the $r$-charges 
used in the orbifolding 
and the preserved fermionic charges 
%on the lattice 
differ from the ones in \citen{Cohen:2003qw}.

We have shown that the calculation of
the partition function 
can be facilitated like as the equation (\ref{partition-func-lat}) 
by the localization theorem
on our lattice theory.
Such the partition function is expressed 
in terms of integrals only on
the $M$ eigenvalues $\phi_{\nv,i}$ of each $\Phi_\nv$.
We can adopt the 
technique analogous to %he one in
\citen{Moore:1998et}
since our lattice model is constructed by the orbifolding of the 
matrix model whose partition function is 
calculated in 
\citen{Moore:1998et}. 
In this calculation on the lattice, we do not need the 
parameter $\e$ and $\tilde{\e}$
in order to regularize the partition function since 
the orbifold projection plays the same role of the $\Omega$-background.
Our remaining task about the calculation of 
partition function
is to perform the
integration over $\Phi$ on the lattice theory.
One can also regard
this facilitating technique as
the 
%existence of 
Nicolai map \cite{Sakai:1983dg}. 

We have also investigated about the observables of 
the topological field theory on the lattice. 
These
are defined as the BRST cohomologies on the lattice obtained by
the
orbifold projection of the BRST cohomologies in the matrix model.
Only the 0-form observables 
${\cal O}_0$ which are the polynomials of the BRST cohomological 
field $\Phi$ can be non-trivial 
on the lattice 
while the observables 
corresponding to 1-form or 2-form 
in continuum theory
must be BRST trivial on the lattice.
It is worthwhile for the studies of topology on the lattice 
to consider how dimensional reduction or
lattice regularization makes such observables trivial. 
One of the reasons 
for the above facts
would be that 
${\cal O}_0$ is independent of the space-time coordinates
while
other operators are defined with the integral over 
homology cycles which become ill-defined in the matrix and lattice model. 
This consideration should gain something to do with
the study of 
the Leibnitz rule on the lattice, 
the way of triangulation of manifolds, 
%and
%one for 
%related 
%with how we should 
and 
the treatment of the 
%De Rham's duality on the lattice, which 
cohomologies and its dual homologies on the lattice.
%relating to the treatment of 
%related with De Rham's duality
%and 
%the use of Stokes's theorem 
%which 
%can 
%becomes 
%ill-defined on the lattice and matrix model.
%To realize 
%because
%we would obtain further understanding about how to treat
Also for the studies of the space-time generation from the 
matrix model (for example \citen{Ishibashi:1996xs}), 
such the consideration would be important.
It would be 
related with 
the consideration 
how the manifolds with well defined topology
%which can define the topology
should be generated from the matrix model
since our model is constructed from matrix model.
Further works on these problems are needed in the future.

In addition we have shown that
${\cal N} = (2,2)$ supersymmetric Yang-Mills theory
on the lattice
can be obtained from the ${\cal N} = (4,4)$ lattice theory
by the truncation of suitable scalar fields. 
But such lattice model is not equivalent to the model described by 
\citen{Cohen:2003xe}
although the continuum limit of the both are the same as each other.

%\underline{Memo}\\
%: need to discuss with this}\\
The problem on the 
``supersymmetry breaking 
moduli fixing'' 
cannot be ignored in our study 
although we have not investigated it in the present paper.
After performing the deconstruction, 
large fluctuations around the vacuum expectation value
may destroy the lattice structure 
as noted in 
%the series of CKKU's papers 
\citen{Cohen:2003qw}-\citen{Kaplan:2002wv}.
An usual way to suppress such fluctuations
is an addition of  supersymmetry breaking mass terms which 
%destroy the
spoil the topological field theoretical property 
on the lattice.
But there is a possibility to suppress such the fluctuations without any supersymmetry
breaking.
Using the modification of BRST symmetry
with the parameter
$\e$ as explained in
the equation (\ref{BRST-def_e})
in Section
\ref{Sec:partition},
we can suppress the fluctuations around the vacuum expectation value
since the modification introduce the mass term
without the breaking of the modified BRST symmetry.
%\\
%\underline{Memo: Need to discuss with this}\\

Our ${\cal N} = (2,2)$ model 
obtained by the truncation
would 
not suffer from the moduli fixing problem
since
there are not the scalar fields which give the large fluctuations 
destroying the lattice structure.
%by the definition of truncation.
So we might be able to 
define lattice theories without the supersymmetry breaking mass term
which spoil the topological field theoretical property after the truncation.
%which do not 
%As one of pending works,

\section*{Acknowledgments}

We would like to thank
T.~Asakawa,
S.~Catterall,
H.~Fukaya,
A.~Hanany,
I.~Kanamori,
H.~Kawai,
T.~Kuroki,
S.~Matsuura,
T.~Onogi,
H.~Pfeiffer,
Y.~Shibusa,
F.~Sugino,
H.~Suzuki,
N.~Yokoi for useful comments and discussions.
We also thank the Yukawa Institute for Theoretical Physics at Kyoto University. Discussions during the YITP workshop YITP-W-06-11 on ``String Theory and Quantum Field Theory'' were useful to complete this work.
KO is supported in part by the Special Postdoctoral Research Fellowship at RIKEN and 21st 
Century COE Program at Tohoku University. 
TT is supported in part as Junior Research Associate at RIKEN.

\appendix

\section{Representation of $\gamma_K$ and $U_f$}
\label{Notation}
We adopt the representation of 
$\gamma_K = - \Gamma_6 \Gamma_K$ $( K = 1 \sim 6)$ 
which enables us to write the matrix model action by the `topological field theory form', 
\be
S_0 = \frac{1}{2g^2}Q \Tr 
\left[ \frac{1}{4}\eta[\Phi,\Phib]
+\vec{\chi}'\cdot(\vec{H}'-i\vec{{\cal E}}')
+\lambda^\mu[A_\mu,\Phib]\right]\label{4-tft-reduced-MM}
\ee
obtained by the dimensional reduction from the 
4 dimensional ${\cal N} = 2$ topological field theory.
Here $Q$ is one of 8 supercharges of the matrix model,
which is not only the BRST charge of the
4 dimensional topological field theory but also 
preserved BRST charge in our lattice model. 
We represent the indices $\mu$ ($\mu= 1,\ldots, 4$) as  4 dimensional
space-time indices and gauge fields are represented by
$A_\mu =( A_1,A_2,A_3,A_4)$.
In the action
(\ref{4-tft-reduced-MM}), 
each fermionic field 
$\eta, \lambda^{\mu},\vec{\chi}'$
are 
obtained from the dimensional reduction of 
the differential forms in 4 dimensional 
topological field theory, that is,
$\eta$ comes from the scalar fermionic field
which corresponds to the killing spinor of the BRST charge,
$\lambda^\mu = (\lambda_1,\lambda_2,\lambda_3,\lambda_4)$ are from
the 4 components of fermionic 1-form, 
and the $\vec{\chi}' = (\chi^{12},\chi^{13},\chi^{14})$
are from 3 components of self-dual
fermionic 2-forms. 
In addition, $\vec{H}' = (H^{12}, H^{13}, H^{14})$ are auxiliary
bosonic fields which are superpartners of $\vec{\chi}'$
written as
$(H^{12}, H^{13}, H^{14})=(Q\chi^{12},Q\chi^{13},Q\chi^{14})$.

In order to rewrite the matrix model action (\ref{reduced MM})
into (\ref{4-tft-reduced-MM}),
we take the 
basis of 8 component complex fermion $\Psi$ as
\be
\Psi^T=(\lambda_2,\lambda_3,\lambda_4,\lambda_1,\chi^{12},\chi^{13},
\chi^{14},\frac{1}{2}\eta),\label{basis}
%=(\lambda,\lambda^\dag,\lambdat,\lambdat^\dag,\chi^\R,\chi^\C,
%{\chi^\C}^\dag,\frac{1}{2}\eta) \cdot U_f^T, 
\ee
with the following representation of 
%becomes the symmetric matrices  
$\gamma_K$, 
\[
\begin{split}
&\gamma_k = -i 
\begin{pmatrix}
0 & \mu_k \\
\mu^{T}_k & 0 
\end{pmatrix}\qquad
(k = 1 \cdots 4),\\
&\gamma_{5} = i\sigma_3 \otimes \mathbf{1}_{4}, \quad \gamma_{6} = \mathbf{1},
\end{split}
\]
where 
\[
\mu_1
= \begin{pmatrix}
-\mathbf{1} & 0 \\
0 & -\sigma_3
\end{pmatrix}
, \,
\mu_2
= \begin{pmatrix}
0 & \sigma_1 \\
-i\sigma_2 & 0
\end{pmatrix}
, \,
\mu_3
= \begin{pmatrix}
0 & -\sigma_3 \\
\mathbf{1} & 0
\end{pmatrix}
, \,
\mu_4
= \begin{pmatrix}
i\sigma_2 & 0 \\
0 & \sigma_1
\end{pmatrix}.
\]

To generate the lattice theory by using the 
``orbifold projection'',
it is convenient to introduce the complex matrix coordinates
obtained by the combination of the above variables
$A_{\mu},\lambda^{\mu},\cdots$, ${\it etc}$. 
We introduce the 
complex bosonic matrix coordinates
%for the later convenience,
%which are like as
defined as
(\ref{complex-boson-matrix}),
%And
and similarly the complexified auxiliary fields 
$H^\R,H^\C,H^{\C\dag}$, which is defined by
\[
H^\R = H^{14}, \quad
H^{\C} = H^{13} -i H^{12}, \quad
H^{\C\dag} = H^{13} +i H^{12},
\]
and their
BRST partners
%and the  
\[
(\lambda,\lambda^\dag,\lambdat,\lambdat^\dag,\chi^\R,\chi^\C,
{\chi^\C}^\dag,\frac{1}{2}\eta). 
\]
This set of the fermions 
%and the original base
is written in terms of the linear combination of 
the basis (\ref{basis})
using the $8 \times 8$ matrix
$U_f$ as
\[
(\lambda,\lambda^\dag,\lambdat,\lambdat^\dag,\chi^\R,\chi^\C,
{\chi^\C}^\dag,\frac{1}{2}\eta) \cdot U_f^T=
(\lambda_2,\lambda_3,\lambda_4,\lambda_1,\chi^{12},\chi^{13},
\chi^{14},\frac{1}{2}\eta),
\]
where $U_f$ is given by
\[
U_f = \begin{pmatrix}
A_f & 0 \\
0 & B_f
\end{pmatrix},
\]
with
\[
A_f = \begin{pmatrix}
  0 & 0 & -\frac{i}{2} & \frac{i}{2} \\
  0 & 0 & -\frac{1}{2} & -\frac{1}{2} \\
  \frac{i}{2} & -\frac{i}{2} &  0 & 0 \\ 
  \frac{1}{2} & \frac{1}{2} &  0 & 0 \\ 
\end{pmatrix}
,
\quad 
B_f = \begin{pmatrix}
  0 & \frac{i}{2} & -\frac{i}{2} & 0\\
  0 & \frac{1}{2} & \frac{1}{2} & 0  \\
  1 & 0 &  0 & 0 \\ 
  0 & 0 &  0 & 1 \\ 
\end{pmatrix}
.
\]
In this paper,
we take the basis of fermion field $\Psi$ as
\[
\Psi^T=(\lambda_2,\lambda_3,\lambda_4,\lambda_1,\chi^{12},\chi^{13},
\chi^{14},\frac{1}{2}\eta)
=(\lambda,\lambda^\dag,\lambdat,\lambdat^\dag,\chi^\R,\chi^\C,
{\chi^\C}^\dag,\frac{1}{2}\eta) \cdot U_f^T,
\]
using the above expression of $\gamma$ matrices.

%\subsubsection{The relation ship with the ${\cal N} = (2,2)$ CKKU model} 

%%%
%%% References
%%%
\providecommand{\href}[2]{#2}\begingroup\raggedright\endgroup


\begin{thebibliography}{10}


\bibitem{Seiberg:1994rs}
N.~Seiberg and E.~Witten,
  {\it Electric - magnetic duality, monopole
  condensation, and confinement in N=2 supersymmetric yang-mills theory},
  \NPB{426,1994,19--52},
  [\href{http://xxx.lanl.gov/abs/hep-th/9407087}{{\tt hep-th/9407087}}].

\bibitem{Seiberg:1994aj}
N.~Seiberg and E.~Witten, {\it Monopoles, duality and chiral symmetry breaking
  in N=2 supersymmetric QCD},  \NPB{431,1994,484--550},
  [\href{http://xxx.lanl.gov/abs/hep-th/9408099}{{\tt hep-th/9408099}}].

\bibitem{Nekrasov:2002qd}
N.~A. Nekrasov, {\it Seiberg-witten prepotential from instanton counting},
  {\em Adv. Theor. Math. Phys.} {\bf 7} (2004) 831--864,
  [\href{http://xxx.lanl.gov/abs/hep-th/0206161}{{\tt hep-th/0206161}}].

\bibitem{Nekrasov:2003af}
N.~A. Nekrasov, {\it Seiberg-witten prepotential from instanton counting},
  \href{http://xxx.lanl.gov/abs/hep-th/0306211}{{\tt hep-th/0306211}}.

\bibitem{Dijkgraaf:2002fc}
R.~Dijkgraaf and C.~Vafa, {\it Matrix models, topological strings, and
  supersymmetric gauge theories},  \NPB{644,2002,3--20},
  [\href{http://xxx.lanl.gov/abs/hep-th/0206255}{{\tt hep-th/0206255}}].

\bibitem{Dijkgraaf:2002vw}
R.~Dijkgraaf and C.~Vafa, {\it On geometry and matrix models},  \NPB{644,2002,21--39},
  [\href{http://xxx.lanl.gov/abs/hep-th/0207106}{{\tt hep-th/0207106}}].

\bibitem{Dijkgraaf:2002dh}
R.~Dijkgraaf and C.~Vafa, {\it A perturbative window into non-perturbative
  physics},  \href{http://xxx.lanl.gov/abs/hep-th/0208048}{{\tt
  hep-th/0208048}}.


%%%%%%%%%%%%%%%%%%%%%%%%%%%%%%%%%%%%%%%%%%%%%%%%%%%%%%%%%%%%%%%%%%%
%
%       ###  analitic
%
%%%%%%%%%%%%%%%%%%%%%%%%%%%%%%%%%%%%%%%%%%%%%%%%%%%%%%%%%%%%%%%%%%%




%%%%%%%%%%%%%%%%%%%%%%%%%%%%%%%%%%%%%%%%%%%%%%%%%%%%%%%%%%%%%%%%%%%
%
%       ###  SUSY -Lat
%
%%%%%%%%%%%%%%%%%%%%%%%%%%%%%%%%%%%%%%%%%%%%%%%%%%%%%%%%%%%%%%%%%%%%%
%%%%%%%%%%%%%%%%%%%%%%%%%% CKKU  %%%%%%%%%%%%%%%%%%%%%%%%%%%%%%%%%%%%%

\bibitem{Cohen:2003qw}
A.~G. Cohen, D.~B. Kaplan, E.~Katz, and M.~Unsal, {\it Supersymmetry on a
  euclidean spacetime lattice. ii: Target theories with eight supercharges},
  \JHEP{12,2003,031},
  [\href{http://xxx.lanl.gov/abs/hep-lat/0307012}{{\tt hep-lat/0307012}}].

\bibitem{Cohen:2003xe}
A.~G. Cohen, D.~B. Kaplan, E.~Katz, and M.~Unsal, {\it Supersymmetry on a
  euclidean spacetime lattice. i: A target theory with four supercharges},
  \JHEP{08,2003,024},
  [\href{http://xxx.lanl.gov/abs/hep-lat/0302017}{{\tt hep-lat/0302017}}].

\bibitem{Kaplan:2002wv}
D.~B. Kaplan, E.~Katz, and M.~Unsal, {\it Supersymmetry on a spatial lattice},
  \JHEP{05,2003,037},
  [\href{http://xxx.lanl.gov/abs/hep-lat/0206019}{{\tt hep-lat/0206019}}];
D.~B. Kaplan and M.~Unsal, {\it A euclidean lattice construction of
  supersymmetric yang- mills theories with sixteen supercharges},  \JHEP{09,2005,042}, [\href{http://xxx.lanl.gov/abs/hep-lat/0503039}{{\tt
  hep-lat/0503039}}];
M.~G. Endres and D.~B. Kaplan, {\it Lattice formulation of (2,2) supersymmetric
  gauge theories with matter fields},
  \href{http://xxx.lanl.gov/abs/hep-lat/0604012}{{\tt hep-lat/0604012}};
M.~Unsal, {\it Twisted supersymmetric gauge theories and orbifold lattices},
  \href{http://xxx.lanl.gov/abs/hep-th/0603046}{{\tt hep-th/0603046}}.

%%%%%%%%%%%%%%%%%%%%%%%%%%%%% polar %%%%%%%%%%%%%%%%%%%%%%%%%%%%%%

\bibitem{Unsal:2005yh}
M.~Unsal, {\it Compact gauge fields for supersymmetric lattices},  \JHEP{11,2005,013}, [\href{http://xxx.lanl.gov/abs/hep-lat/0504016}{{\tt
  hep-lat/0504016}}].

\bibitem{Onogi:2005cz}
T.~Onogi and T.~Takimi, {\it Perturbative study of the supersymmetric lattice
  theory from matrix model},  \PRD{72,2005,074504},
  [\href{http://xxx.lanl.gov/abs/hep-lat/0506014}{{\tt hep-lat/0506014}}].


%\cite{Catterall:2003wd}
\bibitem{Catterall:2003wd}
  S.~Catterall,
  {\it Lattice supersymmetry and topological field theory
  },
  JHEP {\bf 0305}, 038 (2003)
  [\href{http://xxx.lanl.gov/abs/hep-lat/0301028}{{\tt hep-lat/0301028}}];
  %%CITATION = HEP-LAT 0301028;%%
S.~Catterall, {\it A geometrical approach to N = 2 super Yang-Mills theory on
  the two dimensional lattice},  \JHEP{11,2004,006},
  [\href{http://xxx.lanl.gov/abs/hep-lat/0410052}{{\tt hep-lat/0410052}}];

%%%%%%%%%%%%%%%%%%%%%%%%%%%%% Sugino %%%%%%%%%%%%%%%%%%%%%%%%%%%%%%%

\bibitem{Sugino:2003yb}
F.~Sugino, {\it A lattice formulation of super yang-mills theories with exact
  supersymmetry},  \JHEP{01,2004,015},
  [\href{http://xxx.lanl.gov/abs/hep-lat/0311021}{{\tt hep-lat/0311021}}];
F.~Sugino, {\it Super yang-mills theories on the two-dimensional lattice with
  exact supersymmetry},  \JHEP{03,2004,067},
  [\href{http://xxx.lanl.gov/abs/hep-lat/0401017}{{\tt hep-lat/0401017}}];
F.~Sugino, {\it A lattice formulation of super yang-mills theories with exact
  supersymmetry},  {\em Nucl. Phys. Proc. Suppl.} {\bf 140} (2005) 763--765,
  [\href{http://xxx.lanl.gov/abs/hep-lat/0409036}{{\tt hep-lat/0409036}}];
F.~Sugino, {\it Various super yang-mills theories with exact supersymmetry on
  the lattice},  \JHEP{01,2005,016},
  [\href{http://xxx.lanl.gov/abs/hep-lat/0410035}{{\tt hep-lat/0410035}}];
F.~Sugino, {\it Two-dimensional compact n = (2,2) lattice super yang-mills
  theory with exact supersymmetry},  \PLB{635,2006,218--224},
   [\href{http://xxx.lanl.gov/abs/hep-lat/0601024}{{\tt
  hep-lat/0601024}}].

%%%%%%%%%%%%%%%%%%%%%%%% Giedt %%%%%%%%%%%%%%%%%%%%%%%%%%%%%%%% 

\bibitem{Giedt:2003ve}
J.~Giedt, {\it Non-positive fermion determinants in lattice supersymmetry},
  \NPB{668,2003,138--150},
  [\href{http://xxx.lanl.gov/abs/hep-lat/0304006}{{\tt hep-lat/0304006}}];
J.~Giedt, {\it The fermion determinant in (4,4) 2d lattice super-yang- mills},
  \NPB{674,2003,259--270},
  [\href{http://xxx.lanl.gov/abs/hep-lat/0307024}{{\tt hep-lat/0307024}}];
J.~Giedt, {\it Deconstruction, 2d lattice yang-mills, and the dynamical lattice
  spacing},  \href{http://xxx.lanl.gov/abs/hep-lat/0312020}{{\tt
  hep-lat/0312020}};
J.~Giedt, {\it Deconstruction, 2d lattice super-yang-mills, and the dynamical
  lattice spacing},  \href{http://xxx.lanl.gov/abs/hep-lat/0405021}{{\tt
  hep-lat/0405021}};
J.~Giedt, {\it Deconstruction and other approaches to supersymmetric lattice
  field theories},  \IJMP{A21,2006,3039--3094},
  [\href{http://xxx.lanl.gov/abs/hep-lat/0602007}{{\tt hep-lat/0602007}}];
J.~Giedt, {\it Quiver lattice supersymmetric matter, d1/d5 branes and
  ads(3)/cft(2)},  \href{http://xxx.lanl.gov/abs/hep-lat/0605004}{{\tt
  hep-lat/0605004}}.

%%%%%%%%%%%%%%%%%%%%%%% beyond tree %%%%%%%%%%%%%%%%%%%%%%%%%%%%%%%%

\bibitem{Elliott:2005bd}
J.~W. Elliott and G.~D. Moore, {\it Three dimensional N = 2 supersymmetry on
  the lattice},  {\em PoS} {\bf LAT2005} (2006) 245,
  [\href{http://xxx.lanl.gov/abs/hep-lat/0509032}{{\tt hep-lat/0509032}}].

\bibitem{Feo:2004kx}
A.~Feo, {\it Supersymmetry on the lattice},  {\em Nucl. Phys. Proc. Suppl.}
  {\bf 119} (2003) 198--209,
  [\href{http://xxx.lanl.gov/abs/hep-lat/0210015}{{\tt hep-lat/0210015}}];
A.~Feo, {\it The supersymmetric ward-takahashi identity in 1-loop lattice
  perturbation theory. i: General procedure},  \PRD{70,2004,054504},
  [\href{http://xxx.lanl.gov/abs/hep-lat/0305020}{{\tt
  hep-lat/0305020}}];
A.~Feo, P.~Merlatti, and F.~Sannino, {\it Information on the super yang-mills
  spectrum},  {\em Phys. Rev.} {\bf D70} (2004) 096004,
  [\href{http://xxx.lanl.gov/abs/hep-th/0408214}{{\tt hep-th/0408214}}];
A.~Feo, {\it Predictions and recent results in susy on the lattice},  {\em Mod.
  Phys. Lett.} {\bf A19} (2004) 2387--2402,
  [\href{http://xxx.lanl.gov/abs/hep-lat/0410012}{{\tt hep-lat/0410012}}].

\bibitem{Catterall}
S.~Catterall, {\it Lattice formulation of n = 4 super yang-mills theory},  {\em
  JHEP} {\bf 06} (2005) 027,
  [\href{http://xxx.lanl.gov/abs/hep-lat/0503036}{{\tt hep-lat/0503036}}].
%%%%%%%%%%%%%%%%%%%%%%%%%%%% From earlier %%%%%%%%%%%%%%%%%%%%%%%%%%



\bibitem{Elitzur:1982vh}
S.~Elitzur, E.~Rabinovici, and A.~Schwimmer, {\it Supersymmetric models on the
  lattice},  \PLB{119,1982,165};
I.~Ichinose, {\it Supersymmetric lattice gauge theory},  \PLB{122,1983,68};
R.~Nakayama and Y.~Okada, {\it Supercurrent anomaly in lattice gauge theory},
  \PLB{134,1984,241};
G.~Curci and G.~Veneziano, {\it Supersymmetry and the lattice: A
  reconciliation?},  \NPB{292,1987,555}.

\bibitem{Nishimura:1997vg}
J.~Nishimura, {\it Four-dimensional n = 1 supersymmetric yang-mills theory on
  the lattice without fine-tuning},  \PLB{406,1997,215--218},
  [\href{http://xxx.lanl.gov/abs/hep-lat/9701013}{{\tt
  hep-lat/9701013}}];
N.~Maru and J.~Nishimura, {\it Lattice formulation of supersymmetric yang-mills
  theories without fine-tuning},  \IJMP{A13,1998,2841--2856},
   [\href{http://xxx.lanl.gov/abs/hep-th/9705152}{{\tt
  hep-th/9705152}}].

\bibitem{Taniguchi:1999fc}
Y.~Taniguchi, {\it One loop calculation of susy ward-takahashi identity on
  lattice with wilson fermion},  {\em Chin. J. Phys.} {\bf 38} (2000) 655--662,
  [\href{http://xxx.lanl.gov/abs/hep-lat/9906026}{{\tt hep-lat/9906026}}].

\bibitem{Kaplan:1999jn}
D.~B. Kaplan and M.~Schmaltz, {\it Supersymmetric yang-mills theories from
  domain wall fermions},  {\em Chin. J. Phys.} {\bf 38} (2000) 543--550,
  [\href{http://xxx.lanl.gov/abs/hep-lat/0002030}{{\tt hep-lat/0002030}}];
G.~T. Fleming, J.~B. Kogut, and P.~M. Vranas, {\it Super yang-mills on the
  lattice with domain wall fermions},  \PRD{64,2001,034510},
   [\href{http://xxx.lanl.gov/abs/hep-lat/0008009}{{\tt
  hep-lat/0008009}}].



\bibitem{Montvay:2001kw}
I.~Montvay {\em et~al.}, {\it Numerical simulation of supersymmetric yang-mills
  theory}, Prepared for NIC Symposium 2001, Julich, Germany, 5-6 Dec 2001;
I.~Montvay, {\it Supersymmetric yang-mills theory on the lattice},  \IJMP{A17,2002,2377--2412},
  [\href{http://xxx.lanl.gov/abs/hep-lat/0112007}{{\tt hep-lat/0112007}}].

\bibitem{Farchioni:2001wx}
{\bf DESY-Munster-Roma} Collaboration, F.~Farchioni {\em et~al.}, {\it The
  supersymmetric ward identities on the lattice},  {\em Eur. Phys. J.} {\bf
  C23} (2002) 719--734, [\href{http://xxx.lanl.gov/abs/hep-lat/0111008}{{\tt
  hep-lat/0111008}}].



\bibitem{Itoh:2002nq}
K.~Itoh, M.~Kato, H.~Sawanaka, H.~So, and N.~Ukita, {\it Towards the super
  yang-mills theory on the lattice},  \PTP{108,2002,363--374},
   [\href{http://xxx.lanl.gov/abs/hep-lat/0112052}{{\tt
  hep-lat/0112052}}];
K.~Itoh, M.~Kato, H.~Sawanaka, H.~So, and N.~Ukita, {\it Novel approach to
  super yang-mills theory on lattice: Exact fermionic symmetry and 'ichimatsu'
  pattern},  \JHEP{02,2003,033},
  [\href{http://xxx.lanl.gov/abs/hep-lat/0210049}{{\tt hep-lat/0210049}}].


\bibitem{Harada:2003bs}
M.~Harada and S.~Pinsky, {\it N = (1,1) super yang-mills on a (2+1) dimensional
  transverse lattice with one exact supersymmetry},  \PLB{567,2003,277--287},
   [\href{http://xxx.lanl.gov/abs/hep-lat/0303027}{{\tt
  hep-lat/0303027}}];
M.~Harada and S.~Pinsky, {\it N = 1 super yang-mills on a (3+1) dimensional
  transverse lattice with one exact supersymmetry},  \PRD{71,2005,065013},
   [\href{http://xxx.lanl.gov/abs/hep-lat/0411024}{{\tt
  hep-lat/0411024}}].


\bibitem{DAdda:2005zk}
A.~D'Adda, I.~Kanamori, N.~Kawamoto, and K.~Nagata, {\it Exact extended
  supersymmetry on a lattice: Twisted n = 2 super yang-mills in two
  dimensions},  \PLB{633,2006,645--652},
  [\href{http://xxx.lanl.gov/abs/hep-lat/0507029}{{\tt hep-lat/0507029}}];
F.~Bruckmann, S.~Catterall, and M.~de~Kok, {\it A critique of the link approach
  to exact lattice supersymmetry},
  \href{http://xxx.lanl.gov/abs/hep-lat/0611001}{{\tt hep-lat/0611001}}.



\bibitem{Suzuki:2005dx}
H.~Suzuki and Y.~Taniguchi, {\it Two-dimensional n = (2,2) super yang-mills
  theory on the lattice via dimensional reduction},  \JHEP{10,2005,082},
   [\href{http://xxx.lanl.gov/abs/hep-lat/0507019}{{\tt hep-lat/0507019}}];
H.~Fukaya, I.~Kanamori, H.~Suzuki, M.~Hayakawa, and T.~Takimi, {\it Note on
  massless bosonic states in two-dimensional field theories},
  \href{http://xxx.lanl.gov/abs/hep-th/0609049}{{\tt hep-th/0609049}}.



%%%%%%%%%%%%%%
%%%%%%%%%%%%%%

%\cite{Fukuma:1992hy}
\bibitem{Fukuma:1992hy}
  M.~Fukuma, S.~Hosono and H.~Kawai,
  %``Lattice topological field theory in two-dimensions,''
  Commun.\ Math.\ Phys.\  {\bf 161}, 157 (1994)
  [arXiv:hep-th/9212154];
  %%CITATION = HEP-TH 9212154;%%
  S.~w.~Chung, M.~Fukuma and A.~D.~Shapere,
  %``Structure of topological lattice field theories in three-dimensions,''
  Int.\ J.\ Mod.\ Phys.\ A {\bf 9}, 1305 (1994)
  [arXiv:hep-th/9305080].
  %%CITATION = HEP-TH 9305080;%%

\bibitem{Pfeiffer}
 A.~D.~Lauda and H.~Pfeiffer,
 ``Open-closed strings: Two-dimensional extended TQFTs and Frobenius  algebras,''
  \href{http://www.citebase.org/abstract?id=oai:arXiv.org:math/0510664}{{\tt math.AT/0510664}};
%  A.~D.~Lauda and H.~Pfeiffer,
 ``State sum construction of two-dimensional open-closed Topological Quantum Field Theories,''
  \href{http://www.citebase.org/abstract?id=oai:arXiv.org:math/0602047}{{\tt math.QA/0602047}};






%%%%%%%%%%%%%%%%%%%%%%%%%%%%%%%%%%%%%%%%%%%%%%%%%%%%%%%%%%%%%%%%%%%
%
%      #orbifold and others
%
%%%%%%%%%%%%%%%%%%%%%%%%%%%%%%%%%%%%%%%%%%%%%%%%%%%%%%%%%%%%%%%
\bibitem{Douglas:1996sw}
M.~R. Douglas and G.~W. Moore, {\it D-branes, quivers, and ale instantons},
  \href{http://xxx.lanl.gov/abs/hep-th/9603167}{{\tt hep-th/9603167}}.

\bibitem{Kachru:1998ys}
S.~Kachru and E.~Silverstein, {\it 4d conformal theories and strings on
  orbifolds},  \PRL{80,1998,4855--4858},
  [\href{http://xxx.lanl.gov/abs/hep-th/9802183}{{\tt hep-th/9802183}}].

\bibitem{Arkani-Hamed:2001ca}
N.~Arkani-Hamed, A.~G. Cohen, and H.~Georgi, {\it (de)constructing dimensions},
   \PRL{86,2001,4757--4761},
  [\href{http://xxx.lanl.gov/abs/hep-th/0104005}{{\tt hep-th/0104005}}].

\bibitem{Arkani-Hamed:2001ie}
N.~Arkani-Hamed, A.~G. Cohen, D.~B. Kaplan, A.~Karch, and L.~Motl, {\it
  Deconstructing (2,0) and little string theories},  \JHEP{01,2003,083},
   [\href{http://xxx.lanl.gov/abs/hep-th/0110146}{{\tt hep-th/0110146}}].

\bibitem{vanNieuwenhuizen:1996tv}
P.~van Nieuwenhuizen and A.~Waldron, {\it On euclidean spinors and wick
  rotations},  \PLB{389,1996,29--36},
  [\href{http://xxx.lanl.gov/abs/hep-th/9608174}{{\tt hep-th/9608174}}].

\bibitem{Hirano:1997ai}
S.~Hirano and M.~Kato, {\it Topological matrix model},  \PTP{98,1997,1371--1384},
  [\href{http://xxx.lanl.gov/abs/hep-th/9708039}{{\tt hep-th/9708039}}].

\bibitem{Moore:1998et}
G.~W. Moore, N.~Nekrasov, and S.~Shatashvili, {\it D-particle bound states and
  generalized instantons},  \CMP{209,2000,77--95},
  [\href{http://xxx.lanl.gov/abs/hep-th/9803265}{{\tt hep-th/9803265}}].

\bibitem{Hanany:1997tb}
A.~Hanany and A.~Zaffaroni, {\it On the realization of chiral four-dimensional
  gauge theories using branes},  \JHEP{05,1998,001},
  [\href{http://xxx.lanl.gov/abs/hep-th/9801134}{{\tt hep-th/9801134}}].

\bibitem{Hanany:1998it}
A.~Hanany and A.~M. Uranga, {\it Brane boxes and branes on singularities},
  \JHEP{05,1998,013},
  [\href{http://xxx.lanl.gov/abs/hep-th/9805139}{{\tt hep-th/9805139}}].

\bibitem{Bershadsky:1995qy}
M.~Bershadsky, C.~Vafa, and V.~Sadov, {\it D-branes and topological field
  theories},  \NPB{463,1996,420--434},
  [\href{http://xxx.lanl.gov/abs/hep-th/9511222}{{\tt hep-th/9511222}}].

\bibitem{Ooguri:2002gx}
H.~Ooguri and C.~Vafa, {\it Worldsheet derivation of a large N duality},  \NPB{641,2002,3--34},
  [\href{http://xxx.lanl.gov/abs/hep-th/0205297}{{\tt hep-th/0205297}}].

\bibitem{Moore:1997dj}
G.~W. Moore, N.~Nekrasov, and S.~Shatashvili, {\it Integrating over higgs
  branches},  \CMP{209,2000,97--121},
  [\href{http://xxx.lanl.gov/abs/hep-th/9712241}{{\tt hep-th/9712241}}].

\bibitem{Bruzzo:2002xf}
U.~Bruzzo, F.~Fucito, J.~F. Morales, and A.~Tanzini, {\it Multi-instanton
  calculus and equivariant cohomology},  \JHEP{05,2003,054},
  [\href{http://xxx.lanl.gov/abs/hep-th/0211108}{{\tt hep-th/0211108}}].

\bibitem{Fucito:2004gi}
F.~Fucito, J.~F. Morales, and R.~Poghossian, {\it Instantons on quivers and
  orientifolds},  \JHEP{10,2004,037},
  [\href{http://xxx.lanl.gov/abs/hep-th/0408090}{{\tt hep-th/0408090}}].

\bibitem{Sakai:1983dg}
H.~Nicolai, {\it On a new characterization of supersymmetric theories. 2},
  CERN-TH-2811;
H.~Nicolai, {\it Supersymmetry and functional integration measures}, \NPB{176,1980,419--428};
N.~Sakai and M.~Sakamoto, {\it Lattice supersymmetry and the nicolai mapping},
  \NPB{229,1983,173};
Y.~Kikukawa and Y.~Nakayama, {\it Nicolai mapping vs. exact chiral symmetry on
  the lattice},  \PRD{66,2002,094508},
  [\href{http://xxx.lanl.gov/abs/hep-lat/0207013}{{\tt hep-lat/0207013}}].

\bibitem{Witten:1988ze}
E.~Witten, {\it Topological quantum field theory},  \CMP{117,1988,353};
E.~Witten, {\it Introduction to cohomological field theories},  \IJMP{A6,1991,2775--2792};
M.~Blau and G.~Thompson, {\it Lectures on 2-d gauge theories: Topological
  aspects and path integral techniques},
  \href{http://xxx.lanl.gov/abs/hep-th/9310144}{{\tt hep-th/9310144}}.


\bibitem{Ishibashi:1996xs}
N.~Ishibashi, H.~Kawai, Y.~Kitazawa, and A.~Tsuchiya, {\it A large-N reduced
  model as superstring},  \NPB{498,1997,467--491},
  [\href{http://xxx.lanl.gov/abs/hep-th/9612115}{{\tt hep-th/9612115}}].

\end{thebibliography}
\end{document}